\documentclass{aa}  
\usepackage[utf8]{inputenc}
\usepackage{graphicx}
\usepackage{amsmath}	
\usepackage{amssymb}	
\usepackage{multirow}
\usepackage{relsize}
\usepackage{txfonts}
\usepackage[normalem]{ulem}
\usepackage{makecell}
\usepackage{hyperref}
\hypersetup{
    colorlinks=true,
    breaklinks=true,
    allcolors=blue
}

\newcommand{\mhi}{M_{\rm{HI}}}
\newcommand{\kms}{km\,s$^{-1}$}

\newcommand{\Msun}{M$_\odot$}

\newcommand{\HI}{{\sc H\,i }}

\newcommand{\Mdm}{M_{\rm{DM}}(<20\,\rm kpc)}
\graphicspath{{figure/}}

\begin{document} 
    
    \title{Uncovering the dark matter distribution by combining stellar kinematics and integrated \HI spectra: Method validation}
    \titlerunning{}
    \authorrunning{Lei et al.}  
    \subtitle{}
    \author{
    Yu Lei$^{1,2}$, Meng Yang$^{1}$\thanks{myang@shao.ac.cn}, Ling Zhu$^{1}$\thanks{lzhu@shao.ac.cn} 
    }
    \institute{
Shanghai Astronomical Observatory, Chinese Academy of Sciences, 80 Nandan Road, Shanghai 200030, China
\and
School of Astronomy and Space Sciences, University of Chinese Academy of Sciences, No. 19A Yuquan Road, Beijing 100049, China
}

   \date{Received; accepted}
 
   \abstract
   {
   We determined the dark matter (DM) distribution in galaxies by jointly modelling stellar kinematics from integral field unit (IFU) observations and the gaseous kinematics encoded in a single integrated \HI spectrum. The stellar kinematics are described by a triaxial orbit-superposition Schwarzschild model, while the \HI gas is described by an idealised disc model; both are governed by the same gravitational potential. The potential comprises the stellar mass, a generalised Navarro–Frenk–White (gNFW) DM halo, and a central black hole. We validated the method on 58 simulated galaxies generated from the TNG50 cosmological simulation. For each galaxy, we created two versions of mock data with azimuthal angles viewed side-on and end-on, thus yielding 116 mock observations in total. Our model recovers the total mass, stellar mass, and DM mass profiles within the data range; the median DM mass of the 58 simulated galaxies is recovered with a relative systematic bias smaller than 20\% across all radii from 2--20~kpc. The statistical uncertainties on the DM masses within 5 kpc remain similar to those found with the model constrained by IFU data only. In contrast, the relative uncertainty on the DM mass in the outer regions decreases when the \HI spectrum is included; at 20 kpc, it drops markedly, from about 85\% to roughly 30\%. The DM density slope defined explicitly in the gNFW model is systematically underestimated and thus does not yield a reliable quantity from observations using our approach. Instead, we introduce density slopes evaluated between 2 and 20 kpc, which are statistically well recovered for both the total mass and the DM mass. We demonstrate the reliability of this method in uncovering the DM distribution and emphasise its promise for application to large samples of observed galaxies.

}


   \maketitle

\section{Introduction}

Uncovering the dark matter (DM) distribution in galaxies is fundamental to understanding the co-evolution of galaxies and their host DM haloes. One of the primary techniques for assessing the DM distribution in nearby galaxies is through kinematics. Over the past two decades, integral field unit (IFU) spectrograph surveys, such as ATLAS$^{\rm{3D}}$ \citep{Cappellari2011}, CALIFA \citep{Sanchez2012}, SAMI \citep{Croom2012}, and MaNGA \citep{Bundy2015}, have provided stellar kinematic data for tens of thousands of nearby galaxies. 
Stellar kinematics are complicated by the diverse motion of stars. By creating dynamical models to fit the stellar kinematic data, the underlying mass distribution within the data coverage can be well uncovered. 
Using these data, researchers have been able to measure the DM content in a substantial sample of these galaxies, employing different dynamical modelling approaches, including the Jeans Anisotropic Modelling (JAM) method \citep[e.g.][]{2013MNRAS.432.1709C,2024MNRAS.527..706Z}, the orbit-superposition Schwarzschild model \citep[e.g.][]{Jin2020MNRAS.491.1690J,2022ApJ...930..153S}, and the action-based distribution function method \citep[e.g.][]{Gherghinescu2024MNRAS.533.4393G}. Compared with simpler approaches such as JAM, the Schwarzschild method makes fewer assumptions about the functional form of the distribution function, enabling a more flexible reconstruction of the stellar orbital structure.
In particular, the Schwarzschild method has been validated via tests with mock data created from different simulations \citep{zhu2018MNRAS.473.3000Z,jin2019MNRAS.486.4753J}. The circular velocity derived from this method has also been shown to be consistent with that observed from the molecular gas (CO) for CALIFA galaxies \citep{Leung2018MNRAS.477..254L}.

However, observations from large IFU surveys typically cover only the luminous regions of galaxies, approximately within 1--2 effective radii ($R_{\rm e}$). For most galaxies, the gravitational potential in the central regions is dominated by baryons, with DM contributing only a minor fraction within the spatial range of IFU data. The limited spatial coverage of the data potentially causes large uncertainties in the measurement of DM fractions. We need kinematic tracers that extend to larger radii, where DM becomes dominant, to accurately determine the DM mass in nearby galaxies.

Extended stellar kinematics, obtained from IFU with a larger field of view~\citep[e.g.][]{2023A&A...675A.143C,2024MNRAS.528.5295Y}, long-slit spectroscopy~\citep[e.g.][]{2007MNRAS.382..657T,2009MNRAS.400.1665W,2010ApJ...716..370F}, and multi-object spectroscopy~\citep[e.g.][]{2015A&A...574A..93L}, have been used to probe the faint galaxy outskirts and constrain their mass distributions. However, the long exposure times required have limited the number of such datasets. Alternatively, discrete tracers, such as globular clusters (GCs) and planetary nebulae (PNe), can extend dynamical measurements to even larger radii~\citep[e.g.][]{2011MNRAS.411.2035N,2016MNRAS.460.3838A,2017MNRAS.468.3949A,2024A&A...685A.132D}. However, sufficient numbers of these tracers are usually available only in the most massive early-type galaxies.

Neutral hydrogen (\HI) gaseous discs are ubiquitous in late-type galaxies and are often found in field early-type galaxies~\citep{2006MNRAS.371..157M,2009A&A...498..407G,2012MNRAS.422.1835S}. For systems in dynamical equilibrium, \HI gas moving in nearly circular orbits provides a well-established probe of the gravitational potential. \HI rotation curves have been used for decades to measure DM distributions, mainly in late-type galaxies, but also in early-type galaxies~\citep[e.g.][]{1985ApJ...295..305V,2006PhDT.........1N,2016AJ....152..157L}. More recently, combining stellar IFU data with spatially resolved \HI velocity fields has been proposed as a powerful approach to constrain DM profiles across the full radial range---from galaxy centres to the outskirts~\citep{2020MNRAS.491.4221Y,2024MNRAS.528.5295Y}, yet such samples remain limited. The largest sample of galaxies with resolved \HI rotation curves in the literature is the SPARC sample \citep{Lelli2016AJ....152..157L}, which consists mostly of late-type galaxies within 20~Mpc. However, the SPARC galaxies have very limited overlap with large IFU surveys. Radio telescopes typically lack the spatial resolution required to resolve \HI kinematics except for very nearby galaxies, whereas IFU instruments offer high spatial resolution but have a limited field of view, which prevents them from targeting these local systems.

The \HI integrated spectra, with no spatial resolution, are much more cost-effective and are already available for large galaxy samples. 
Historically, the \HI line width has been widely used as a dynamical proxy in galaxy scaling relations, most notably the Tully--Fisher relation and its baryonic extension~\citep[e.g.][]{1977A&A....54..661T,2000ApJ...533L..99M,2019MNRAS.484.3267L}, and more recently to infer DM halo properties \citep{2023MNRAS.526.5861Y}.
In addition, combining the rotational velocity derived from the \HI line width with the sizes of gaseous discs provides an efficient way to estimate the dynamical mass of galaxies within a certain aperture~\citep[e.g.][]{2018MNRAS.478.1611K,2020ApJ...898..102Y, 2026A&A...706A..64Y}. 
Beyond the line width alone, the full shape of the integrated \HI profile contains additional dynamical information. \citet{2021arXiv210504570P} presented a preliminary study using the full distribution of \HI velocity profiles to constrain halo properties. 
Building on this, forward modelling of the integrated \HI spectrum profile can further improve the constraints~\citep{Yasin2023MNRAS.525.5066Y, 2026MNRAS.tmp..558Y}. For galaxies with \HI profiles that are not strongly asymmetric, the rotation curves of galaxies were shown to be reasonably well recovered from integrated \HI spectra, whereas the rotation velocities at the relatively outer regions, near the maximum of $r\Sigma_{\rm HI}$, are best constrained~\citep{2026MNRAS.tmp..558Y}.

There is a large overlap between the galaxies from IFU surveys and those with observed integrated \HI spectra. These two datasets are highly complementary for constraining the DM distribution. Integral field unit surveys provide stellar kinematics within the inner 1-2 $R_{\rm e}$, thereby strongly constraining the mass distribution in the central regions. By contrast, integrated \HI spectra are more effective at capturing the total rotational velocity and thus the DM distribution at galaxy outer regions. In this work, we validated the method of uncovering the DM distribution by combining MaNGA-like IFU stellar kinematics with \HI integrated spectra using mock galaxies generated from cosmological simulations. Different IFU observations can vary significantly in spatial coverage, data quality, and related characteristics and thereby differ in how effectively they can constrain the DM mass distribution. Throughout this paper, we refer to IFU data with similar properties from the MaNGA survey. We adopted the DYNAMITE package \citep{2020ascl.soft11007J}---a version of the Schwarzschild code originally from \citet{vdB2008MNRAS.385..647V} and well tested \citep{Leung2018MNRAS.477..254L,zhu2018MNRAS.473.3000Z,jin2019MNRAS.486.4753J}---for modelling the stellar kinematics as well as a thin disc model for the \HI gas.

The paper is organised as follows. Section~\ref{sec:sample} introduces the simulated galaxies used to construct the mock observations. Section~\ref{sec:method} describes the method for combining stellar IFU and integrated \HI data to infer the DM distributions. Section~\ref{sec:results} presents the main results, and Sect.~\ref{sec:discussion} provides a detailed discussion. We summarise our findings in Sect.~\ref{sec:summary}.

\section{Data and sample}
\label{sec:sample} 

A large fraction of MaNGA galaxies have \HI spectra observed with the Green Bank Telescope~\citep{2021MNRAS.503.1345S}, which are publicly available as the HI-MaNGA sample~\citep{2019MNRAS.488.3396M}. These data provide exactly the type of IFU stellar kinematics and single-dish \HI spectra that our method is designed to combine. To validate our method under realistic observational conditions, we therefore generated a set of 58 simulated galaxies constructed from cosmological simulations to match the properties and data quality of these observations.

We adopted the cosmological hydrodynamical simulation TNG50~\citep{2019MNRAS.490.3196P,2019MNRAS.490.3234N} to create mock galaxies. TNG50 encompasses a cubic space with each side measuring $51.7\,\mathrm{Mpc}$, simulating $2160^3$ DM particles at a resolution of $4.5\times10^5 M_\odot$, as well as an equivalent number of gas particles at a resolution of $8.5\times10^4 M_\odot$. We used snapshot 99, which corresponds to a redshift value of $z = 0$.

\subsection{Sample of mock galaxies}

The sample of mock galaxies used to validate the method was primarily chosen based on the quality of the mock data.
We computed the total stellar mass and \HI mass for each galaxy by summing over its stellar and gas particles within the corresponding friends-of-friends (FoF) subhalo. This yielded 718 TNG50 galaxies with $M_{\star}>10^{10}\,$\Msun\ and $\mhi > 10^{8.5}\,$\Msun\,.
Integrated \HI\ spectra and stellar-kinematic maps were then constructed as described in Sects.~\ref{subsec:mock_stellar} and \ref{subsec:mock_hi}.
We visually inspected the \HI spectra and selected 321 galaxies with clear double-horned profiles. Double-horned profiles provide us with a realistic way of eliminating galaxies with highly irregular \HI distributions when there is no spatially resolved information available. Although a regular \HI disc can also result in single-peaked profiles \citep{2026MNRAS.tmp..558Y}, such systems are also excluded from our sample. We further excluded 21 systems showing counter-rotating discs in the stellar kinematics, as such systems can exhibit multi-component line-of-sight velocity distributions (LOSVDs) that require a different dynamical-modelling setup. Thus, we do not consider here~\citep{2024MNRAS.528.2643B}.
From the remaining 300 galaxies, we randomly selected 58 simulated galaxies as our final sample. The sample consists of 30 unbarred and 28 barred galaxies, adopting the bar classifications of \citet{2022MNRAS.515.1524Z}, which are based on a Fourier analysis of the stellar surface density distribution.

The selected 58 simulated galaxies span a wide range of DM distributions, and their DM mass fractions are generally representative of TNG50 galaxies in this mass range. 
As a reference, we derived the ground-truth mass properties of the simulated galaxies directly from particle counts. The stellar, DM, and dynamical masses within specific radii were calculated by summing the masses of stars, DM, and all particles within those radii.

\subsection{Projection and viewing angles}

The stellar component of a subhalo is generally triaxial and can be approximated as an ellipsoid with three principal axes (major, intermediate, and minor). We defined these axes using the reduced inertia tensor method of \citet{2006MNRAS.367.1781A}. In brief, we computed the reduced inertia tensor of the stellar particles within a fiducial aperture of 40~kpc,
\begin{equation}
I_{ij}=\sum_n \,\frac{x_{n,i}\,x_{n,j}}{r_n^2},
\end{equation}
where $x_{n,i}$ is the $i$th coordinate of the $n$th particle and $r_n$ is the ellipsoidal radius defined using the current estimate of the principal axes. We iteratively diagonalised $I_{ij}$ and updated $r_n$ until the principal-axis directions converge. The eigenvectors of the converged tensor define the principal-axis frame, which we adopted as the reference coordinate system for our subsequent projections. In this frame, the minor axis is aligned with the $z$-axis, and the major axis is aligned with the $x$-axis. For barred galaxies, we verified that the bar major axis typically aligns with the x-axis in this principal-axis frame under our convention.

We used the inclination and azimuthal angles to project each galaxy onto the 2D observational plane. The inclination angle $\theta$ is defined as the angle between the galaxy minor axis and the line-of-sight direction, such that $\theta=0^\circ$ corresponds to a face-on view and $\theta=90^\circ$ to an edge-on view. The azimuthal angle $\phi$ is defined in the disc plane and is implemented by rotating the system about the $z$-axis prior to projection. With this definition, $\phi= 0^\circ$ corresponds to a side-on view (line of sight perpendicular to the major axis/bar), whereas $\phi=90^\circ$ corresponds to an end-on view (line of sight parallel to the major axis/bar).

The central velocity dispersion in barred galaxies increases as the azimuthal viewing angle changes from side-on to end-on. This behaviour has been reported in simulations \citep[][]{2015MNRAS.450.2514I} and in recent observations \citep{2025A&A...700A.237F}. In this paper, we refer to this behaviour as a bar-like kinematic structure, which is also present in TNG50 (see Fig.~\ref{fig:bar_ang_mock}). Because our dynamical models do not account for such azimuth-dependent non-axisymmetric motions, this effect can potentially bias the inferred mass distribution.

This bar-like kinematic structure is common in our simulated galaxies, including in some galaxies not classified as barred by \citet{2022MNRAS.515.1524Z}. We therefore created two versions of mock data for each simulated galaxy with different azimuthal viewing angles: one viewed side-on, with $\phi$ randomly chosen from $0^\circ<\phi<45^\circ$, and another viewed end-on, with ($45^\circ<\phi<90^\circ$). We randomly chose the inclination angle for each galaxy from $35^\circ<\theta < 90^\circ$ and kept it the same for the two versions of mock data. Face-on galaxies with $\theta < 35^\circ$, which represent only about 20\% of typical observed samples, were not considered because dynamical mass measurements become difficult at such low inclinations: the rotation velocity is poorly constrained and carries large uncertainties. In total, this yielded 116 mock galaxy observations. For each mock observation, we placed the galaxy at a distance of 100~Mpc. Each mock galaxy observation includes the stellar IFU data as well as the integrated \HI spectrum and is analysed as an independent dataset.

\subsection{Mock stellar kinematics}
\label{subsec:mock_stellar}

We created maps of stellar mass surface density, stellar velocity, and velocity dispersion across the 2D observational plane.
The stellar mass surface density map was generated by counting the mass of stellar particles in a grid with a spaxel size of $0.25\times0.25~\rm{arcsec}^2$, following the method described in \citet{2025A&A...700A.249J}.
The stellar velocity and velocity dispersion data were derived as follows. We constructed the LOSVD of all stellar particles on a grid covering a $15\times15~\rm{arcsec}^2$ field of view with a spaxel size of $0.5\times0.5~\rm{arcsec}^2$. The LOSVD in each spaxel was fitted with a Gauss--Hermite series to derive the mean velocity $V$, the velocity dispersion $\sigma$, and the higher-order moments $h_3$ and $h_4$ from which only the velocity and dispersion maps were retained for subsequent analysis. The measurement uncertainties were assigned following the logarithmic function presented by~\citet{2015ApJ...802L...3T} and calibrated from CALIFA data~\citep{2013A&A...549A..87H}. We then added random Gaussian noise to the data with a standard deviation of 0.5 times these uncertainties to both the velocity and dispersion maps. Finally, we applied the Voronoi binning~\citep{2003MNRAS.342..345C} to obtain 250 spatial bins across the field.

\subsection{Mock \HI integrated spectra}
\label{subsec:mock_hi}

We computed the neutral gas fraction of TNG50 gas particles using the \texttt{Hdecompose} package\footnote{\url{https://github.com/kyleaoman/Hdecompose}}.
The calculation follows the multiphase interstellar medium (ISM) model described in~\citet{2003MNRAS.339..289S}.
The molecular gas fraction was then calculated using the $\mathrm{H_2}$ model of~\citet{2011ApJ...728...88G} with the `volumetric' method introduced by~\citet{2015MNRAS.452.3815L} and~\citet{2018ApJS..238...33D}, as implemented in the \texttt{Dirty-AstroPy} package\footnote{\url{https://github.com/arhstevens/Dirty-AstroPy}}.
Finally, the \HI gas mass was derived by first subtracting the molecular component from the total neutral gas mass and then removing the helium contribution.

The \HI spectrum was generated using a modified version of the \texttt{MARTINI} package\footnote{\url{https://github.com/kyleaoman/martini}}~\citep{2019MNRAS.482..821O,2024JOSS....9.6860O}.
We modified the code to produce integrated \HI spectra directly, as the standard \texttt{MARTINI} pipeline was designed to construct full data cubes. 
In our implementation, the \HI emission was modelled as the sum of contributions from the gas particles, each represented by a Gaussian profile centred on its line-of-sight velocity, and the line of sight is identical to that adopted for the mock stellar-kinematic maps. The width of each Gaussian profile was set by thermal broadening and was computed as $\sigma_{\rm{th}}=\sqrt{{k_\mathrm{B}T_{g}}/{m_p}}$, where $k_{\mathrm{B}}$ is Boltzmann's constant, $T_{g}$ is the gas temperature, and $m_p$ is the proton mass. 
Only gas particles located within the 540-arcsec beam were included when constructing the \HI spectrum, and the emission was binned into channels of width $1.4$~\kms. Gaussian noise with an amplitude of 2~mJy was then added. Subsequently, the data were smoothed using a boxcar kernel with a width of four channels, followed by Hanning smoothing to suppress channel-to-channel noise fluctuations. 
This procedure yields an effective velocity resolution of about 10~\kms, consistent with the HI-MaNGA observational setup of~\citet{2019MNRAS.488.3396M}.

\begin{figure*}
\centering\includegraphics[width=0.9\linewidth]{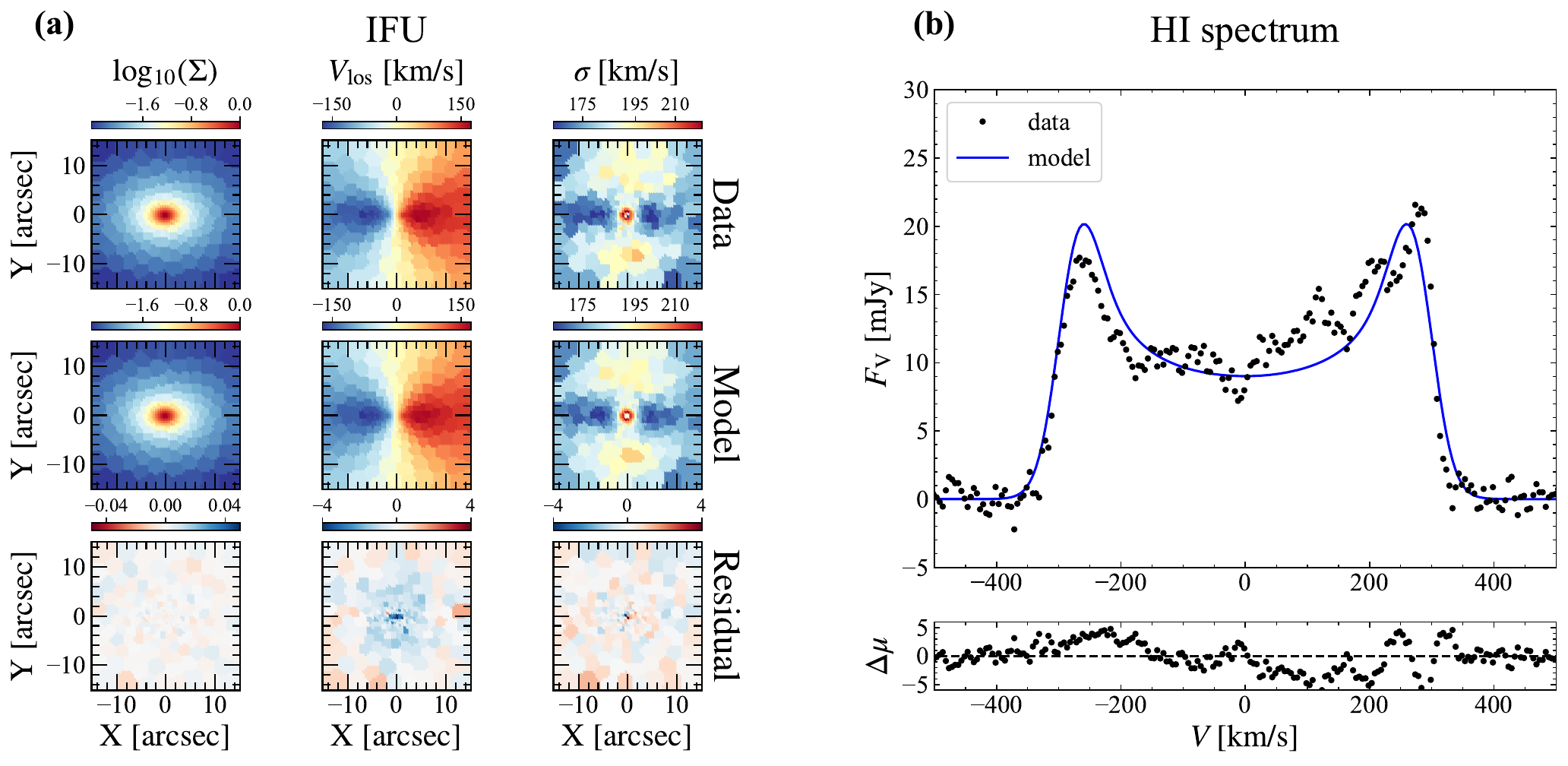}
\caption{
Illustration of dynamical modelling for a mock TNG50 galaxy (TNG50 368843) combining stellar kinematics and integrated \HI spectrum.
Panel (a): Best-fitting model of the stellar kinematics. From left to right: Surface density, velocity, and velocity dispersion. From top to bottom: Data, model, and relative residuals defined as (data-model)/error.
Panel (b): Best-fitting model of the \HI spectrum. Top panel: Data (black points) and the best-fitting model (solid blue line). Bottom panel: Residuals (data-model)/$\sigma_{\rm{RMS}}$, where $\sigma_{\rm{RMS}}$ is the RMS noise of the \HI spectrum.}
\label{fig:model}
\end{figure*}

\begin{figure}
\centering\includegraphics[width=0.4\textwidth]{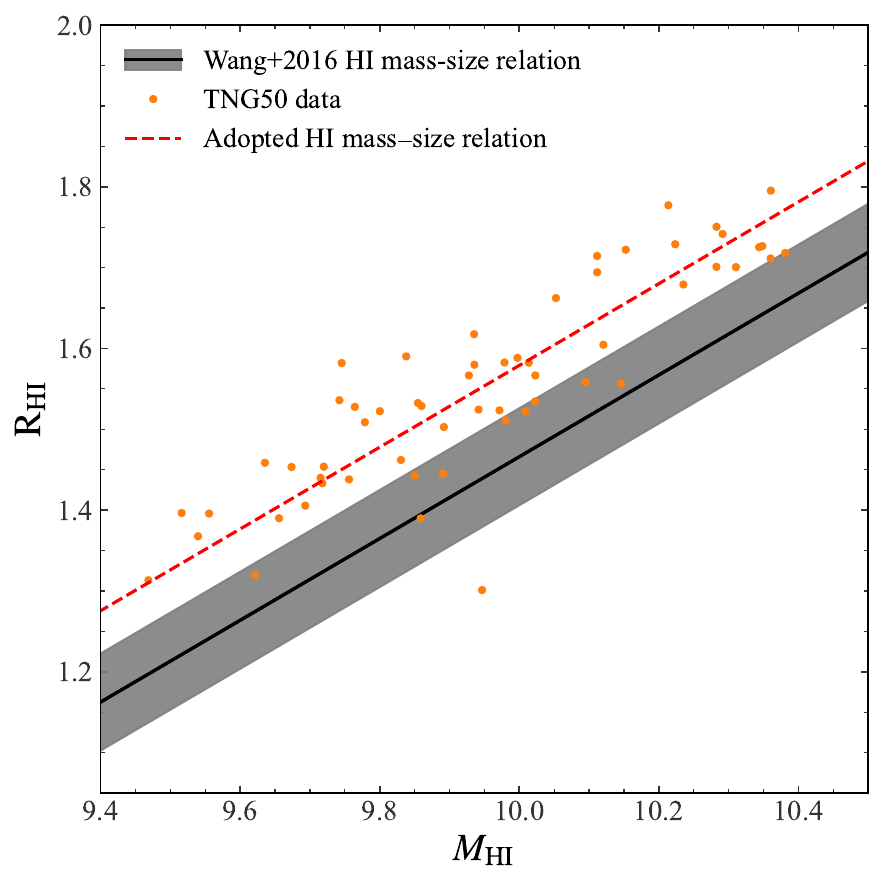}
\caption{\HI mass--size relation. The blue points show our mock galaxies created from TNG50. The solid black line and the grey-shaded region show the relation reported by \citet{2016MNRAS.460.2143W}. The dashed orange line shows our fit to the \HI mass--size relation for the TNG50 mock galaxies, obtained with the slope fixed to the observed value.
} 
\label{fig:hi_mass_size}
\end{figure}

\begin{figure}
\centering\includegraphics[width=0.4\textwidth]{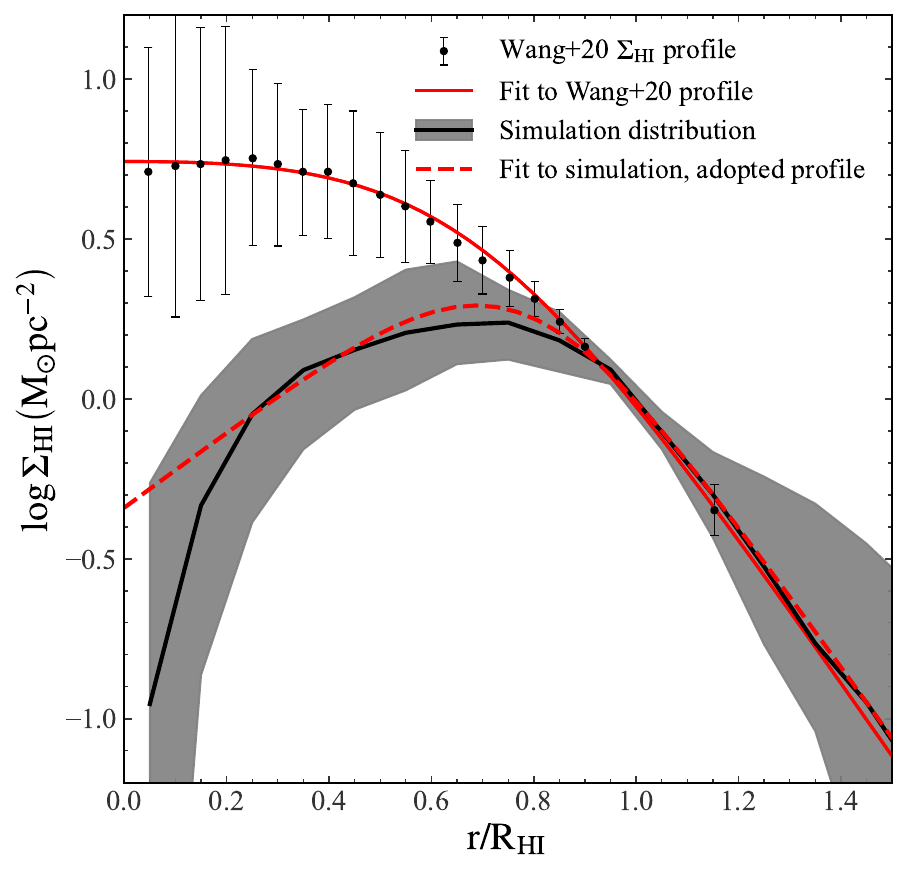}
\caption{\HI surface density profile as a function of normalised radius.
The black points with error bars show the median and scatter of the \HI surface density profiles of 168 nearby galaxies from the observations of \citet{2020ApJ...890...63W}. 
The solid red curve shows the profile fitted to the observed median profile. 
The solid black curve and the grey-shaded region show the median and scatter of the \HI surface-density profiles of the TNG50 mock galaxies.
The dashed red curve shows the profile fitted to the median \HI surface density profile of the TNG50 mock galaxies, which was adopted for constructing the \HI disc model in this work.
The observed profile was adopted in the alternative modelling scenario discussed in Appendix~\ref{app:obs_hi_density}.
} 
\label{fig:hi_density_profile}
\end{figure}

\begin{figure*}
\centering\includegraphics[width=0.8\linewidth]{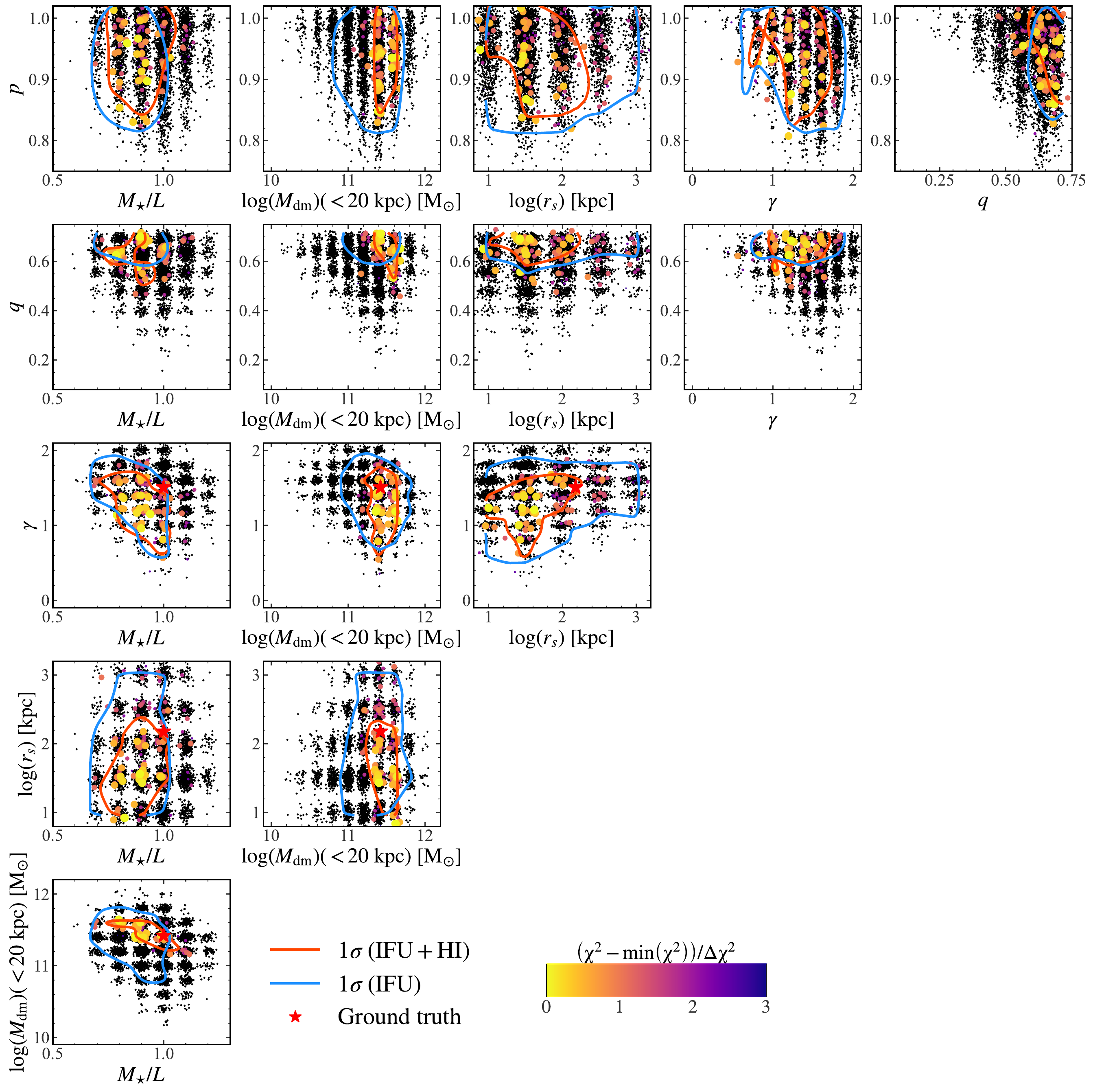}
\caption{Parameter grid explored by the model constrained by the stellar kinematics and integrated \HI spectrum.
We explore six free parameters: the intrinsic axis ratios of the stellar component $p$ and $q$, the stellar mass-to-light ratio $M_{\star}/L$, the DM mass within 20 kpc $\Mdm$, the DM scale radius $r_s$, and the inner slope $\gamma$. The parameters are explored with fixed step sizes; we dithered the points in the figure to make all models visible.
Each point is colour-coded by its $(\chi^2-\min(\chi^2))/\Delta{\chi2}$ value, as indicated by the colour bar. 
The black points denote models outside the $3\sigma$ confidence level. 
The solid orange and blue contours show the $1\sigma$ regions determined by the marginalised posteriors on the 2D projected space, for the joint IFU and \HI constraints (orange) and the IFU-only constraints (blue), respectively. 
The red star marks the ground truth. The constraint on DM mass is significantly improved by combining IFU and HI.} 
\label{fig:grid}
\end{figure*}

\begin{figure*}
\centering\includegraphics[width=0.85\linewidth]{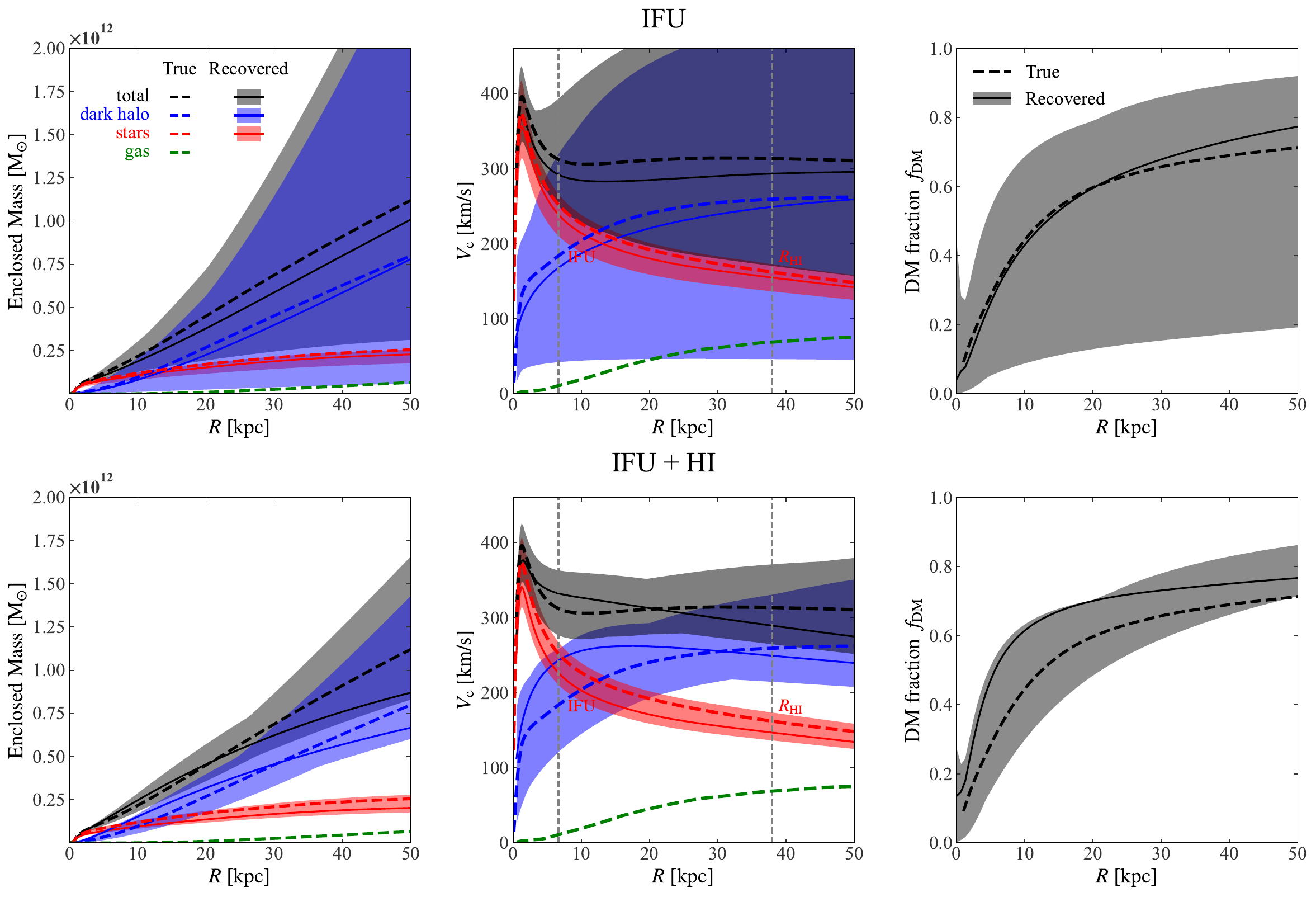}
\caption{Comparison of enclosed mass profiles resulting from the joint IFU and \HI constraints (top rows) and the IFU-only constraints (bottom rows).
Left panels: Enclosed mass profiles for the total (black), dark halo (blue), stellar (red), and gas (green) components. 
The dashed lines indicate the ground truth, while the solid lines with shaded regions show the median model result with its $1\sigma$ uncertainties.
Middle panels: Recovery of the rotation curve. The dashed lines represent the true rotation curve of total mass (black) and those contributed by the stellar (red), DM (blue), and gas (green) mass. The solid lines denote the total (black), stellar (red), and DM (blue) contributions recovered by our dynamical model, with the shaded regions marking the $1\sigma$ uncertainties. The two vertical dashed lines mark the data coverage limits of kinematic map and the \HI radius inferred from the \HI mass--size relation, respectively. 
Right panels: DM fractions $f_{\rm DM}(R)$, with the dashed line showing the ground truth and the solid line with the shaded region representing the median profile with its $1\sigma$ confidence level. 
}
\label{fig:enclosed_mass}
\end{figure*}

\section{Methods}
\label{sec:method}

This section outlines our method for recovering the mass distribution from the mock observations. We first describe the construction of the gravitational potential, the orbit-superposition modelling, and the \HI disc model, and then present the joint analysis that combines stellar kinematics with the integrated \HI constraints.

\subsection{Gravitational potential}
The gravitational potential was constructed from a stellar component, a DM halo, and a central black hole. We adopted the same potential for both the orbit-superposition models and the \HI disc models. 

We described the projected stellar mass density using the 2D multi-Gaussian expansion (MGE) method~\citep{1994A&A...285..723E,2002MNRAS.333..400C}, which decomposes the surface mass density into a set of Gaussian components:
\begin{equation}
S(x',y') = \sum_{i} S_i(x',y') =\sum_{i} \left(\frac{M_i}{2\pi \sigma_i^2 {q_i}}\exp\left[-\frac{1}{2\sigma_i^2}\left(x'^2+\frac{y'^2}{q_i^2}\right)\right]\right),
\end{equation}
where $(x',y')$ are the 2D coordinates aligned along the major and minor axes and $S_i(x',y')$ is the mass distribution of each Gaussian with the corresponding amplitude $M_i$, scale length $\sigma_i$, and flattening $q_i$.

The 2D MGE model was then deprojected onto a triaxial MGE model using a set of viewing angles $(\theta, \psi, \phi)$, which are linked to the intrinsic axial ratios $(p, q, u)$ that define the intrinsic 3D shape of the galaxy~\citep{2008MNRAS.385..647V}. The minor-to-major axis ratio $q$ and the intermediate-to-major axis ratio $p$ were treated as free parameters to allow for triaxiality, while $u$ was fixed to $0.9999$ to limit the model's degrees of freedom. This is a reasonable choice for our sample, which consists mostly disc galaxies without any highly triaxial early-type galaxies.

The DM halo is described with a generalised Navarro–Frenk–White (gNFW) profile~\citep{1996ApJ...462..563N,1996MNRAS.278..488Z}:
\begin{equation}
    \rho_{\mathrm{DM}}(r)=\frac{\rho_0}{\left(\frac{r}{r_{\mathrm{s}}}\right)^\gamma\left(1+\frac{r}{r_{\mathrm{s}}}\right)^{3-\gamma}},
\end{equation}
where $\rho_0$ is the scale density, $r_s$ is the scale radius, and $\gamma$ is the inner slope of the density profile. For the purpose of efficient parameter sampling, the original parameter set $\left(\rho_0, r_{\mathrm{s}}, \gamma \right)$ was reformulated as $\left(\Mdm, r_{\mathrm{s}}, \gamma \right)$, with $\Mdm$ defined as the DM mass enclosed within 20 kpc.

For each galaxy, we included a central black hole treated as a point mass and implemented with a softening length of $10^{-3}$ arcsec. Because the sphere of influence of the black hole is not resolved by the kinematic data, we fixed the black hole mass to $M_{\mathrm{BH}}=10^6\, \mathrm{M}_{\odot}$, and variations in the assumed $M_{\mathrm{BH}}$ do not affect our results.

\subsection{Orbit-superposition method}

The stellar kinematics were modelled using the Schwarzschild orbit-superposition method~\citep{1979ApJ...232..236S}. We adopted the \texttt{DYNAMITE} package~\citep{2020ascl.soft11007J,2022A&A...667A..51T}, which is a modified version of the triaxial dynamical modelling code presented by~\citet{2008MNRAS.385..647V}. 

For each given set of model parameters for the gravitational potential, we created a model to fit the surface mass density and kinematic maps. The model was constructed by sampling two orbit libraries following~\citet{2008MNRAS.385..647V}. Here, we chose $n_E \times n_\theta \times n_R = 21 \times 15 \times 7$ for each orbit library, resulting in a total of 4,410 orbits for each model. 
We then superposed the stellar orbits to reproduce the stellar mass density, velocity, and velocity dispersion maps. We solved the orbit weights by minimising the residuals between the data and the model:
\begin{equation}
\chi_\mathrm{star}^2 = \chi_\mathrm{s,lum}^2 + \chi_\mathrm{s,kin}^2.
\end{equation}
In practice, the residual of the surface mass density $\chi_\mathrm{s,lum}^2$ is negligible, and the residual of stellar kinematics $\chi_\mathrm{s,kin}^2$ dominates $\chi_\mathrm{star}^2$. In Fig.~\ref{fig:model}(a), we present the best-fitting model for the stellar kinematics of a mock TNG50 galaxy.

\subsection{\HI disc model}
\label{sec:hi_density}

We modelled the \HI distribution as an axisymmetric thin disc tracing the same gravitational potential as the stars. We derived the total \HI mass ($M_{\rm HI}$) of each galaxy from its integrated spectral flux \citep{2018ApJ...861...49H}. Since we lacked spatially resolved measurements of the \HI distribution for our sample, we instead inferred their \HI spatial density distributions using empirical relations established from other datasets.

Observationally, the radial extent of the \HI disc is specified by the characteristic radius $R_{\rm HI}$, which is tightly correlated with the total \HI mass in nearby galaxies through the empirical $M_{\rm HI}$--$R_{\rm HI}$ relation \citep{2016MNRAS.460.2143W}. In addition, nearby galaxies show a nearly universal \HI surface density profile when the radius is normalised by $R_{\rm HI}$ \citep{2014MNRAS.441.2159W,2020ApJ...890...63W}.

As shown in Fig.~\ref{fig:hi_mass_size}, the mocks follow a similar trend to the observations but with $R_{\rm HI}$ larger by about $30\%$ at a given $M_{\rm HI}$. We therefore fit the TNG50 $M_{\rm HI}$--$R_{\rm HI}$ relation with the slope fixed to the observed value from \citet{2016MNRAS.460.2143W}. The resulting fit is shown by the dashed orange line in Fig.~\ref{fig:hi_mass_size}. 

The TNG50 mock galaxies also show rather universal \HI density profiles, although with lower central densities compared to the observed galaxies, as shown in Fig.~\ref{fig:hi_density_profile}. Motivated by this, we describe the normalised radial \HI distribution using the empirical form:
\begin{equation}
\label{eqn:HI_density}
    \Sigma_{\rm{HI}}(r)=\frac{\Sigma_{0}\exp(-r/r_s)}{1+\alpha\exp(-r/r_c)},
\end{equation}
where $r=R/R_{\rm{HI}}$, and the normalisation $\Sigma_{0}$ is determined such that the total integrated \HI mass equals $M_{\rm HI}$. 

By fitting Eq.~\ref{eqn:HI_density} to the median curve of the mock galaxies, we obtain $\alpha = 422$, $r_s = 0.19$, and $r_c = 0.12$.
For comparison, fitting the same functional form to the observed median profile from \citet{2020ApJ...890...63W} gives $\alpha = 38$, $r_s = 0.19$, and $r_c = 0.18$. We adopted the $M_{\rm HI}$--$R_{\rm HI}$ relation obtained from the TNG50 mock galaxies, together with the median curve of TNG50-based normalised \HI surface density profile, to construct the \HI surface density profile for our fiducial gas disc model. 

 We then constructed the \HI integrated spectrum following the method adopted in previous studies~\citep[e.g.][]{2009ApJ...698.1467O,2026MNRAS.tmp..558Y}. The \HI disc was modelled as a series of infinitely thin rings deprojected at a common inclination angle. For each ring, the emission was weighted by the \HI surface density profile, and the LOSVD was specified by $V_c(r)$ and $\sigma_{\mathrm{HI}}$. We then summed the flux-weighted LOSVDs of all rings to generate the total spectrum.

For each model with a given potential,  the rotation curve on the equatorial plane was derived as
\begin{equation}
      V_{c}(R)=\left\langle \sqrt{R\,\frac{\partial \Phi(R,\phi,z)}{\partial
      R}}\right\rangle_{\phi,\,z=0}.
\end{equation}
Triaxiality was allowed for the stellar component in our model. In such a potential, the circular velocity on the equatorial plane formally depends on azimuthal angle $\phi$. We therefore used the azimuthally averaged form above as an effective circular velocity curve. Since most galaxies in our sample are disc-dominated and close to axisymmetric, this approximation is expected to be adequate, although it may be less accurate within the inner few kiloparsec.

In practice, we allowed two additional free parameters in the gaseous disc model: the normalisation $\Sigma_{0}$ of the \HI surface density profile and the intrinsic velocity dispersion $\sigma_{\mathrm{HI}}$. For each model, $\Sigma_{0}$ was allowed to vary within $\pm0.5$ dex from the value determined by the observed $M_{\rm HI}$, while $\sigma_{\mathrm{HI}}$ was allowed to range from 5 to 50~\kms to improve the fit to the integrated spectrum. For most galaxies in our sample, the best-fitting $\Sigma_{0}$ is close to the value determined from $M_{\rm HI}$, and $\sigma_{\mathrm{HI}}$ is approximately 10~\kms, consistent with measurements in nearby \HI galaxies~\citep{2008AJ....136.2563W,2016AJ....151...15M}. 

We compared the \HI spectrum created from the model with the observations. The residual is defined by
\begin{equation}
 \chi^2_\mathrm{gas} = \sum_{i=1}^{N_{\mathrm{bin}}} (\frac{\Psi_\mathrm{obs,i} -\Psi_\mathrm{pred,i} }{\sigma_\mathrm{RMS}}  )^2,
\end{equation}
where $\Psi_{\rm{obs,i}}$ and $\Psi_{\rm{pred,i}}$ are the observed and model-predicted flux in the $i$th velocity bin, respectively. The $\sigma_{\rm{RMS}}$ value represents the noise of the data, measured from a signal-free region of the \HI spectrum. Figure~\ref{fig:model}(b) shows the best-fitting model for the \HI integrated spectrum of the same mock TNG50 galaxy.

We assess the impact of the \HI disc surface density profile on our final results in Appendix~\ref{app:obs_hi_density}, where we discuss constructing the disc model using the observed $M_{\rm HI}$--$R_{\rm HI}$ relation together with the observed median \HI surface-density profile. Since the discrepancy between the TNG50 and observed \HI density profiles is largely confined to the central regions, its effect on the integrated \HI spectrum is minimal, as the line shape is primarily set by gas at larger radii. Consequently, our final results are not significantly altered.

\begin{figure*}
\centering\includegraphics[width=0.9\linewidth]{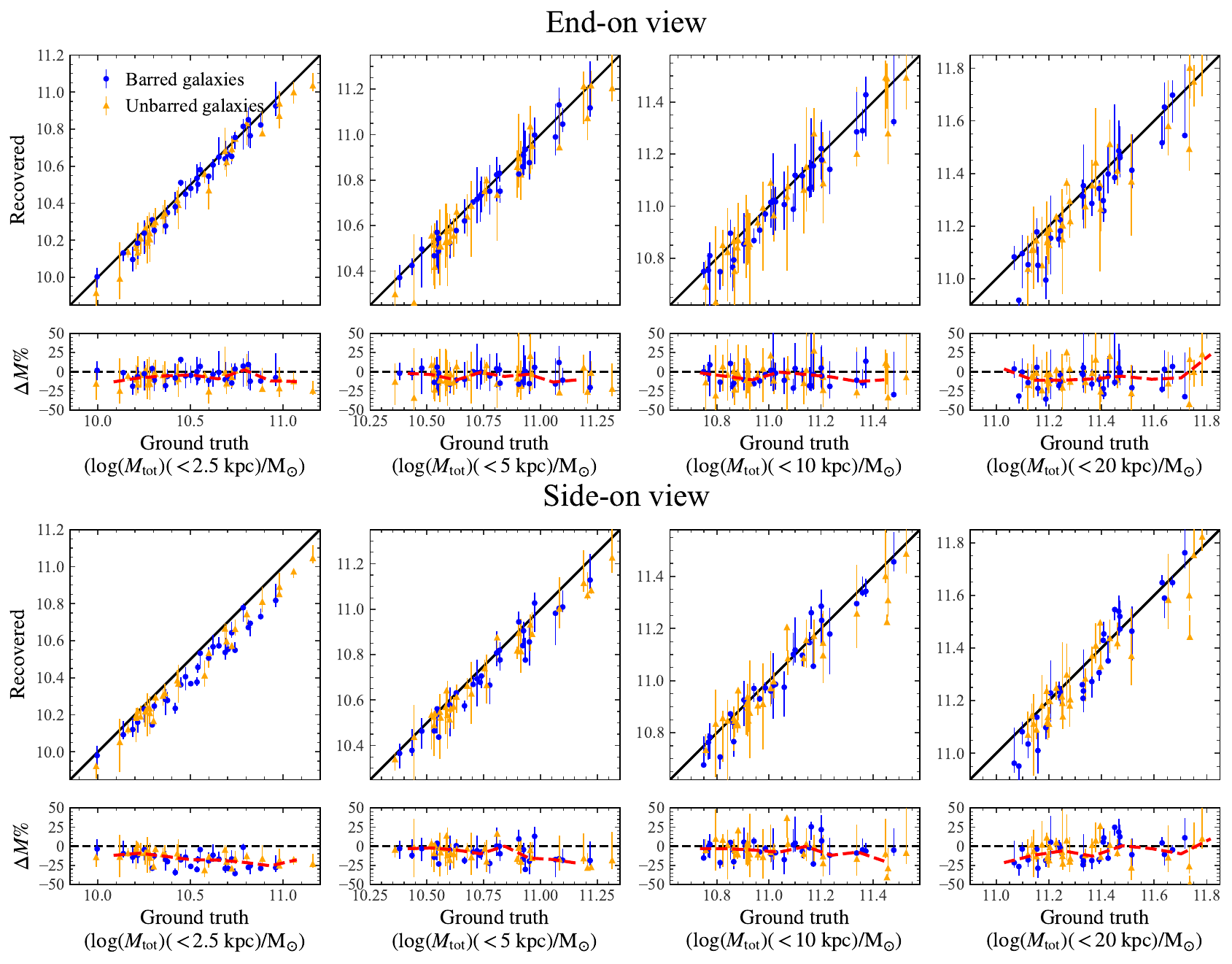}
\caption{Recovery of total mass at different radii for galaxies projected end-on (top) and side-on (bottom), respectively.
We show the one-to-one comparison of the recovered enclosed total mass $M_{\rm tot}$ with the ground truth. From left to right: Enclosed total mass within 2.5, 5, 10, and 20 kpc, respectively. 
Each dot with error bars represents the result of one galaxy, with blue and red denoting barred and unbarred galaxies, respectively.
The diagonal black line marks the one-to-one relation. 
Bottom sub-panels: Fraction of mass deviating from the ground truth $\Delta M/M$, with the dashed lines indicating the median offsets. The total mass at all radii is recovered reasonably well for the end-on galaxies, while there is a systematic bias of $>10\%$ for the side-on galaxies at $r\lesssim 5$ kpc.
}
\label{fig:total_mass_58}
\end{figure*}

\begin{figure}
\centering\includegraphics[width=0.48\textwidth]{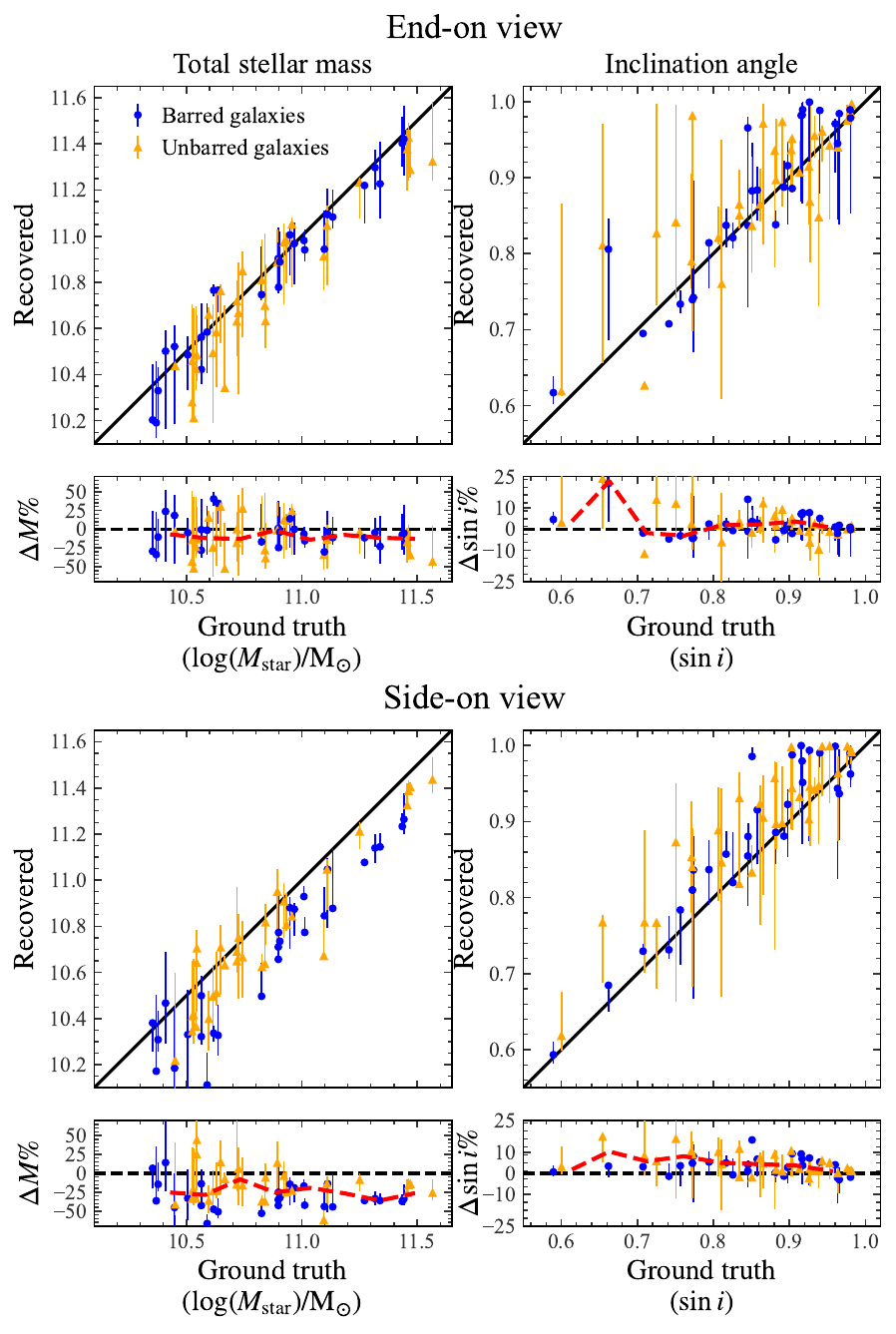}
\caption{Recovery of total stellar mass (left panels) and inclination $\sin i$ (right panels) with the joint IFU and \HI constraints. Top panels: End-on galaxies. Bottom panels: Side-on galaxies. The symbols are the same as in Fig.~\ref{fig:total_mass_58}. } 
\label{fig:stellar_mass_incl_58}
\end{figure}

\begin{figure*}
\centering\includegraphics[width=0.9\linewidth]{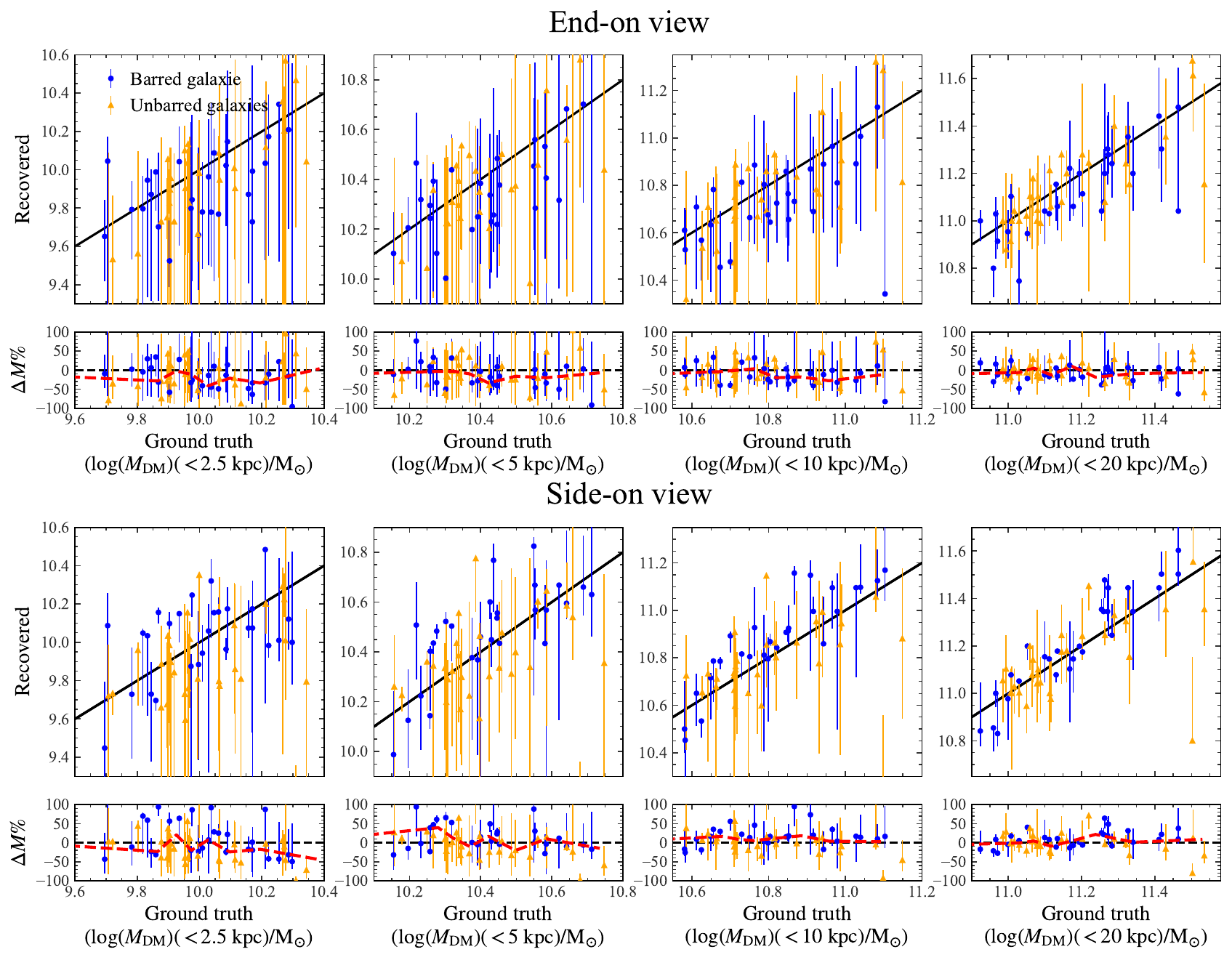}
\caption{Recovery of DM mass at different radii.
The symbols are the same as in Fig.~\ref{fig:total_mass_58}. The uncertainty is large for DM at 2.5 and 5 kpc, where the stellar mass and DM are degenerate and DM mass constitutes a relatively small fraction. The DM masses at all radii are recovered well with no significant systematic bias in both end-on and side-on galaxies.
}
\label{fig:dm_mass_58}
\end{figure*}

\subsection{Joint constraints on the gravitational potential}

We constrained the gravitational potential by combining the fitting of stellar kinematics and \HI spectra. We defined the total residual as 
\begin{equation}
    \chi^2 = \chi^2_{\rm star} + \chi^2_{\rm gas},
\end{equation}
where $\chi^2_{\rm star}$ and $\chi^2_{\rm gas}$ are the residuals of fitting stellar kinematics and \HI spectrum, respectively. 

The gravitational potential model had six free hyper-parameters. Three of them described the DM distribution, the inner density slope $\gamma$, the scale radius $r_s$, and the DM mass enclosed within 20 kpc $M_{\rm DM}(<20 \rm \, kpc)$. The remaining three parameters were the intrinsic shape of the stellar component, $p$, the parameter $q$ related to the three viewing angles, and the stellar mass-to-light ratio, $M_\star/L$. We used an optimised grid searching process \citep{zhu2018MNRAS.473.3000Z} to adjust the free hyper-parameters of the gravitational potential. We started with a model with an initial guess of the hyper-parameters, then we performed an iterative searching process with intervals of 0.25, 0.5, 0.2, 0.08, 0.05, and 0.1 for $\gamma$, $r_s$, $M_{\rm DM}(<20 \rm \, kpc)$, $p$, $q$, and $M_\star/L$. After the previous sampled models were completed, we selected the best-fit models with $\chi^2 - \chi^2_{\rm min} < \chi^2_s$ and sampled the new models around the selected ones by taking two steps in every direction of the parameter grid for each optimal model.
This approach guides the search towards lower $\chi^2$ values within the parameter grid and halts when the minimum $\chi^2$ is identified. The relatively high value of $\chi^2_s$ ($\chi^2_s = \Delta \chi^2_{1\sigma} $) was selected to ensure that all models within at least a $1\sigma$ confidence threshold in the full parameter space were computed before the end of the iteration. Ultimately, we determined the best-fitting models by achieving the minimum $\chi^2$. 

The $1\sigma$ confidence threshold, $\Delta \chi^2_{1\sigma}$, was determined via bootstrapping. We adopted the gravitational potential of the best-fitting model and perturbed the observed velocity and velocity-dispersion maps as well as the integrated \HI spectrum within their respective uncertainties to generate 1000 realisations. In each realisation, we recomputed the model and evaluated the resulting $\chi^2_{\rm kin}$, $\chi^2_{\rm gas}$, and the total $\chi^2$. The standard deviation of these 1000 values of $\chi^2$ was taken as the $1\sigma$ confidence threshold of the model $\Delta \chi^2_{1\sigma} $.

\subsection{Constraints for an example galaxy}
The results of a typical galaxy are shown in Fig.~\ref{fig:grid}. To make all models visible, we dithered the points around the true values of the grid. We determined the $1\sigma$ confidence uncertainty of each parameter by the marginalised posteriors. We considered two versions of the model for each galaxy: one constrained by only the IFU stellar kinematics ($\chi^2_{\rm star}$) and another constrained jointly by IFU and \HI data ($\chi^2$), respectively.  

As illustrated in Fig.~\ref{fig:grid}, the viewing angle parameters $p$ and $q$ and the stellar mass-to-light ratio $M_\star/L$ were constrained in a similarly effective manner and are consistent across the two sets of models. 
Including \HI significantly decreases the uncertainties in the DM distribution parameters, especially regarding the scale radius $r_s$ and the mass $M_{\rm DM} (<20\,\rm kpc)$. The ground truth values of $\gamma$, $r_s$, $M_{\rm DM} (<20\,\rm kpc)$, and $M_\star/L$ are generally recovered by our models within the $1\sigma$ confidence level.

We further derived the enclosed mass profile, DM fraction profile, and rotation curve from our models. We took the best-fitting model as the default result and used the lower and upper limits of the models within the $1\sigma$ confidence threshold as the statistical error.
As shown in Fig.~\ref{fig:enclosed_mass}, the ground truths of these profiles are well recovered by our model constrained by IFU and \HI, within the $1\sigma$ confidence level. The $1\sigma$ relative uncertainty of the enclosed DM mass is about $50\%$ at $r\lesssim 5$ kpc, where stellar mass is dominant, and it is reduced to $30\%$ at $r\sim 20$ kpc. The enclosed DM mass inferred from the IFU-only model has a similar uncertainty at $r\lesssim 5$ kpc, but the uncertainty becomes much larger in the outer regions, reaching $\sim 95\%$ at $r\sim 20$ kpc for this galaxy. Therefore, we focus on the models constrained by IFU and \HI in the following sections.

\begin{table*}[htbp]
\renewcommand{\arraystretch}{1.3}
\small
\centering
\caption{
Relative systematic biases and statistical uncertainties in the recovery of total mass, DM mass, and DM fraction at different radii. 
}
\begin{tabular*}{\textwidth}{@{\extracolsep{\fill}}lllcccc|cccc}
\hline
& & & \multicolumn{4}{c|}{End-on} & \multicolumn{4}{c}{Side-on} \\
\hline
Data & Property &  & $<2.5$ kpc & $<5$ kpc & $<10$ kpc & $<20$ kpc 
     & $<2.5$ kpc & $<5$ kpc & $<10$ kpc & $<20$ kpc \\
\hline

\multirow{6}{*}{IFU+\HI}
& \multirow{2}{*}{Total mass} 
& Bias          
& $-7_{-2}^{+1}\%$ & $-5_{-6}^{+2}\%$ & $-7_{-2}^{+3}\%$ & $-10_{-1}^{+4}\%$ 
& $-15_{-2}^{+1}\%$ & $-9_{-2}^{+3}\%$ & $-5_{-2}^{+0}\%$ & $-8_{-3}^{+4}\%$ \\
& 
& Uncertainty   
& $14\%$ & $21\%$ & $24\%$ & $17\%$ 
& $9\%$ & $12\%$ & $13\%$ & $11\%$ \\

& \multirow{2}{*}{DM mass} 
& Bias          
& $-18_{-8}^{+4}\%$ & $-16_{-3}^{+7}\%$ & $-9_{-4}^{+1}\%$ & $-7_{-5}^{+8}\%$ 
& $-13_{-5}^{+10}\%$ & $2_{-6}^{+6}\%$ & $10_{-3}^{+3}\%$ & $4_{-3}^{+4}\%$ \\
& 
& Uncertainty   
& $68\%$ & $57\%$ & $45\%$ & $30\%$ 
& $49\%$ & $40\%$ & $33\%$ & $20\%$ \\

& \multirow{2}{*}{DM fraction} 
& Bias          
& $-14_{-6}^{+4}\%$ & $-8_{-2}^{+2}\%$ & $-4_{-5}^{+4}\%$ & $3_{-2}^{+3}\%$ 
& $14_{-12}^{+6}\%$ & $18_{-7}^{+4}\%$ & $17_{-3}^{+3}\%$ & $14_{-1}^{+1}\%$ \\
& 
& Uncertainty   
& \makecell[c]{+62\%\\-56\%} 
& \makecell[c]{+40\%\\-49\%} 
& \makecell[c]{+24\%\\-32\%} 
& \makecell[c]{+12\%\\-14\%} 

& \makecell[c]{+26\%\\-50\%} 
& \makecell[c]{+16\%\\-38\%}
& \makecell[c]{+9\%\\-24\%}
& \makecell[c]{+5\%\\-10\%}\\

\hline\hline  
\multirow{6}{*}{IFU only}
& \multirow{2}{*}{Total mass} 
& Bias          
& $-6_{-3}^{+1}\%$ & $-4_{-2}^{+1}\%$ & $-4_{-4}^{+2}\%$ & $-11_{-1}^{+2}\%$
& \multicolumn{4}{c}{--} \\
& 
& Uncertainty   
& $12\%$ & $21\%$ & $34\%$ & $58\%$ 
& \multicolumn{4}{c}{--} \\

& \multirow{2}{*}{DM mass} 
& Bias          
& $-4_{-7}^{+6}\%$ & $-4_{-4}^{+7}\%$ & $-3_{-6}^{+5}\%$ & $-8_{-3}^{+7}\%$
& \multicolumn{4}{c}{--} \\
& 
& Uncertainty   
& $68\%$ & $62\%$ & $66\%$ & $85\%$
& \multicolumn{4}{c}{--} \\

& \multirow{2}{*}{DM fraction} 
& Bias          
&$3_{-8}^{+5}\%$ & $3_{-3}^{+6}\%$ & $6_{-2}^{+2}\%$ & $8_{-1}^{+2}\%$
& \multicolumn{4}{c}{--} \\
& 

& Uncertainty   
& \makecell[c]{+52\%\\-71\%} 
& \makecell[c]{+32\%\\-58\%} 
& \makecell[c]{+24\%\\-44\%} 
& \makecell[c]{+15\%\\-33\%} 
& \multicolumn{4}{c}{--} \\

\hline
\end{tabular*}
\tablefoot{The bias is defined as the median fractional residual over the corresponding galaxy subsample. 
The subscript and superscript on each bias value give the 16th--84th percentile range from 10,000 bootstrap resamplings of the galaxies in that subsample.
The row labelled `Uncertainty' provides the median statistical uncertainty of the individual galaxies. 
The results are shown separately for the models constrained with IFU and \HI and IFU only. 
For the IFU and \HI models, we considered both end-on and side-on projections of the bar-like structures, whereas for the IFU-only case we completed the modelling only for the end-on projection.}
\label{tab:bias_uncertainty}
\end{table*}

\begin{figure}
\centering\includegraphics[width=0.48\textwidth]{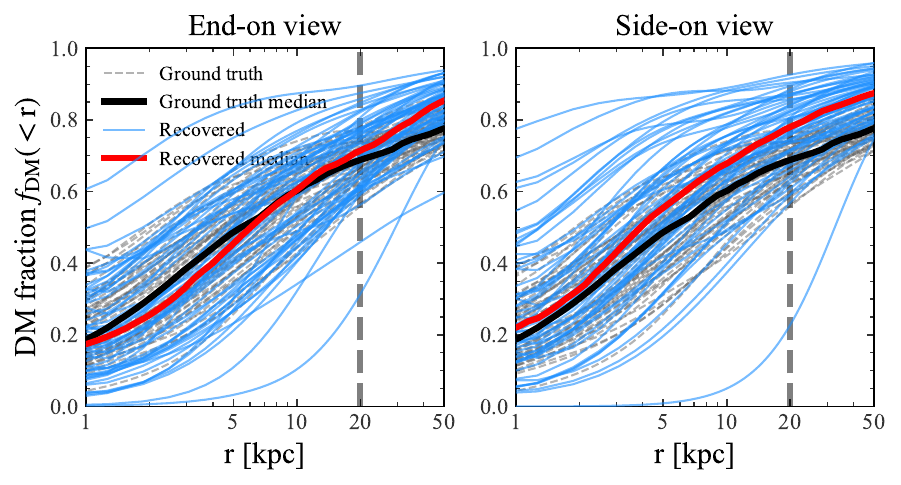}
\caption{Recovery of DM fractions as a function of radius for end-on (left) and side-on (right) galaxies. Each thin grey curve represents the ground truth, and each thin blue curve represents the best-fitting model of one galaxy. The thick black and red curves represent the median of ground truth and the median model recovery of the 58 simulated galaxies.}
\label{fig:dmfrac_curve_58}
\end{figure}

\begin{figure}
\centering\includegraphics[width=0.48\textwidth]{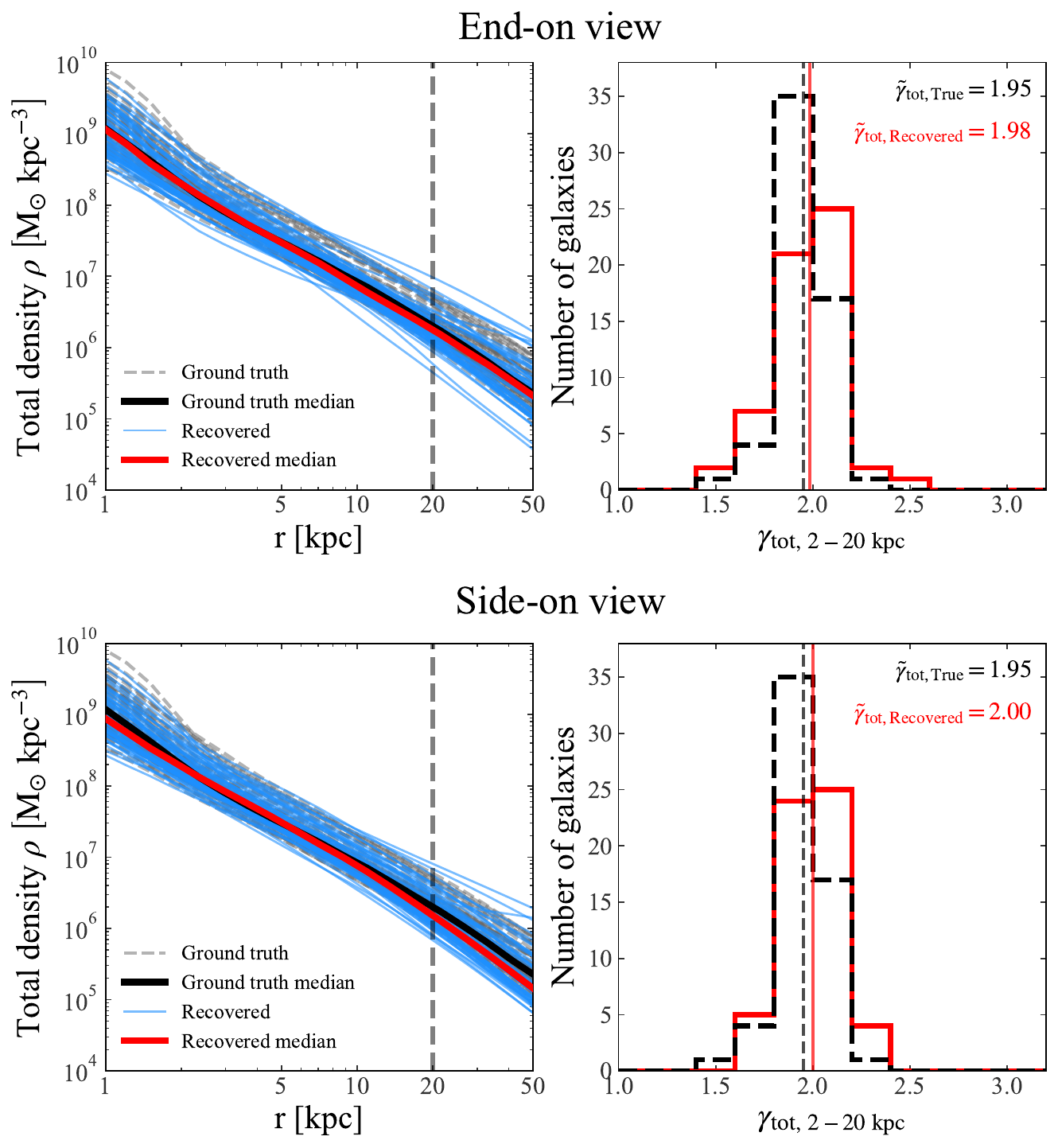}
\caption{Recovery of the total density profiles and the density slope $\gamma_{\rm tot}$ of the total density profile for the end-on (top) and side-on (bottom) galaxies. Left: Ground truth (thin grey curves) and best-fitting model (thin blue curves) of one galaxy. The thick black and red curves represent the median of ground truth and the median model recovery of the 58 simulated galaxies. The total density profiles are well recovered by our model. Right: Distribution of the density slope $\gamma_{\rm tot}$. The dashed black histogram shows the ground truth of $\tilde{\gamma}_{\mathrm{tot,True}}$, obtained by directly fitting to the simulated galaxies. The red histogram shows the recovered values $\tilde{\gamma}_{\mathrm{tot,Recovered}}$ from our modelling. The vertical dashed lines indicate the median values in each case. We recover the $\gamma_{\rm tot}$ values well.
}
\label{fig:tot_density}
\end{figure}

\begin{figure*}[htbp]
\centering
\includegraphics[width=0.9\linewidth]{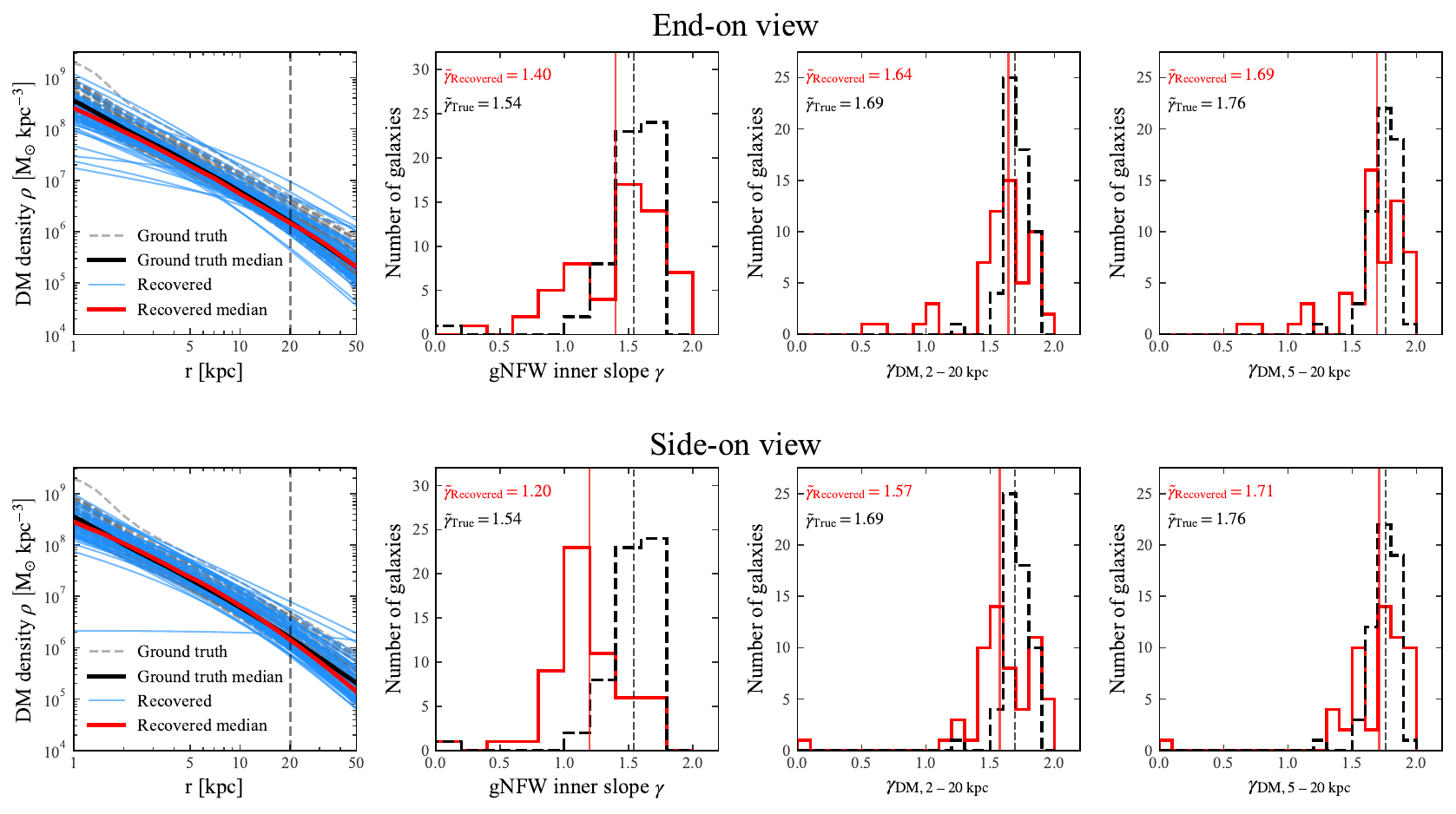}
\caption{
Recovery of the DM density profiles and DM density slopes for the end-on (top) and side-on (bottom) galaxies. First column: DM density profiles of the simulated galaxies, using symbols identical to those in Fig.~\ref{fig:tot_density}. 
Second to fourth columns: Distributions of the DM density slopes, including the gNFW inner slope $\gamma$, and the density slopes $\gamma_{\rm DM,\,2\text{--}20\,{\rm kpc}}$ and $\gamma_{\rm DM,\,5\text{--}20\,{\rm kpc}}$ measured over $2$--$20\,{\rm kpc}$ and $5$--$20\,{\rm kpc}$, respectively.
The dashed black histograms and vertical lines indicate the ground-truth distributions and medians, while the red histograms and vertical lines indicate the recovered distributions and medians. 
The gNFW inner slope is systematically underestimated, especially for the side-on projection, whereas $\gamma_{\rm DM,\,2\text{--}20\,{\rm kpc}}$ and $\gamma_{\rm DM,\,5\text{--}20\,{\rm kpc}}$ are better recovered. 
}
\label{fig:dm_density}
\end{figure*}

\section{Results}
\label{sec:results}

We applied the method to a sample of 116 mock observations created from the 58 simulated galaxies. Here, we demonstrate how the underlying DM distribution is statistically recovered across the full sample by our models constrained by stellar kinematics from IFU observations combined with an integrated \HI spectrum.

\subsection{Recovery of enclosed masses and stellar mass}
\label{ss:totmass}

\subsubsection{Enclosed total, DM, and stellar mass}
We evaluated the total enclosed mass within radii of 2.5, 5, 10, and 20~kpc and compared the recovered masses from our models with the ground-truth values from the simulations. 
The end-on and side-on views are shown separately in Fig.~\ref{fig:total_mass_58}. The ground truths are generally recovered by our model.

For end-on views, the total enclosed mass of each galaxy is well constrained, with typical uncertainties of around $15\%$ in the inner regions ($r<10$ kpc) and increasing to about $20\%$ at $r=20$ kpc. The total mass is slightly underestimated; however, the systematic bias is smaller than $10\%$ across all radii for all galaxies. For side-on views, there is a larger median systematic bias in the inner regions, reaching $-15\%$ at $r=2.5$ kpc. This bias is similar for galaxies classified as barred or unbarred in the Fourier-based catalogue of \citet{2022MNRAS.515.1524Z}, because many nominally unbarred TNG50 systems exhibit bar-like, azimuth-dependent kinematic signatures. This behaviour differs from that found in JAM, which was shown to overestimate the total mass when the bar is viewed end-on and underestimate it when the bar is viewed side-on~\citep{2012MNRAS.424.1495L}.

For each galaxy, we adopted the best-fitting model as the default result and used half the range between the upper and lower limits of the models within the $1\sigma$ confidence threshold as the statistical uncertainty for that galaxy. 
The typical statistical uncertainty was then characterised by taking the median of these uncertainties over the relevant galaxy sample.
We also evaluated the difference between the default results and the ground truths. The systematic bias was defined as the median fractional residual across the corresponding galaxy subsample. Its uncertainty was estimated from 10,000 bootstrap resamples of the galaxy sample. For each resample, we recomputed the median residual and took the 16th and 84th percentiles of the resulting distribution as the uncertainty range.
Detailed systematic biases and statistical uncertainties are summarised in Table~\ref{tab:bias_uncertainty}. In general, the systematic bias of our model is much smaller than the statistical uncertainty.

The total mass comprises both stellar and DM components, which are often degenerate; consequently, the stellar and DM masses generally have larger uncertainties than the total mass. Here, we show the recovery of stellar mass and DM mass, respectively. Figure~\ref{fig:stellar_mass_incl_58} illustrates the recovery of the stellar mass and the inclination angle of the stellar system. The stellar mass recovery is notably accurate, with an individual galaxy uncertainty of around $20\%$; it shows no significant systematic bias ($-2.5\%$) for the galaxies viewed end-on, while there is a bias of $-23\%$ for the galaxies viewed side-on, which is expected as the total mass is underestimated.

The 58 simulated galaxies were randomly projected at inclination angles between $35^\circ$ and $90^\circ$.  In our modelling, the stellar component was treated as triaxial, and the inclination angles were allowed to vary freely. As shown in Fig.~\ref{fig:stellar_mass_incl_58}, the inclination angles of the galaxies are generally well recovered, consistent with the accurate recovery of their total mass profiles.

\subsubsection{Enclosed DM mass and DM fraction}

 The 58 simulated galaxies exhibit a notable variation in their DM distributions. Figure~\ref{fig:dm_mass_58} illustrates the enclosed DM mass within 2.5, 5, 10, and 20 kpc for galaxies viewed end-on and for galaxies viewed side-on. Within 5 kpc, DM contributes only a small fraction of the total mass, leading to relatively large uncertainties in our estimates. The typical uncertainties of DM mass within 2.5 and 5~kpc are $68\%$ and $57\%$ for end-on galaxies, and they decrease to $45\%$ at $r= 10$~kpc and $30\%$ at 20 kpc. There is a similar trend of lower uncertainty for DM in the outer regions for side-on galaxies when combining IFU and \HI data. 
 
 Although there is a relatively large uncertainty for each galaxy, the median systematic bias for the 58 simulated galaxies is below $20\%$ at $2\lesssim r \lesssim 20$~kpc. Specifically, end-on views mildly underestimate the enclosed DM mass over this range, whereas side-on views tend to overestimate it.  This trend is consistent with the larger negative bias in the recovered stellar mass for side-on galaxies ($-23\%$) compared to end-on galaxies ($-2.5\%$), reflecting the degeneracy between the stellar and DM components.

With the enclosed total mass ($M_{\rm{tot}} (<r)$) and DM mass ($M_{\rm DM} (<r)$) profiles derived from the model, we further calculated the DM fraction profile as $f_{\rm DM}(<r) = M_{\rm DM} (<r) / M_{\rm tot} (<r)$. Figure~\ref{fig:dmfrac_curve_58} shows the DM fraction profiles of all 58 simulated galaxies, comparing their ground truth with model recoveries.

Overall, the DM fraction profiles are well recovered by our model. Across the radial range $2 \lesssim r \lesssim 20$ kpc, the median DM fraction of the end-on galaxies is well reproduced, with a median systematic bias below 15\% at all radii. For side-on galaxies, the median DM fraction is overestimated; however, the bias remains below 20\%. 

This overestimate of the DM fraction in the side-on galaxies is mainly driven by the underestimation of the stellar mass. 
In these systems, the side-on stellar kinematics show lower central velocity dispersions than the end-on views, as illustrated in Appendix~\ref{app:azimuth_kinematics}. 
To reproduce this lower dispersion, the dynamical model constrained by the stellar kinematics favour a shallower central gravitational potential. This is mainly achieved by lowering the stellar mass, which dominates the inner regions. The underestimate in the central stellar mass propagates into an underestimate of the total stellar mass, as we use a global scaling parameter for stellar mass.
The outer \HI kinematics then compensate for the underestimated stellar contribution by assigning more mass to the dark halo, leading to an overestimation of the DM mass and hence the DM fraction at larger radii. 

A one-to-one comparison between the true and recovered DM fractions within several radial apertures is presented in Appendix Figs.~\ref{fig:dm_fraction_end_on} and \ref{fig:dm_fraction_side_on}. 
The accurate recovery of the enclosed mass profiles also leads to well-recovered rotation curves, as shown in Appendix Fig.~\ref{fig:rot_curve_58}.

We adopted the best-fitting model as our default result, which differs from the median value of all models within the $1\sigma$ threshold. We then estimated the lower and upper $1\sigma$ errors on the DM fractions separately and find that, in most cases, the lower $1\sigma$ uncertainty is larger than the upper one, as shown in Table~\ref{tab:bias_uncertainty}. The systematic biases are significantly smaller than the statistical uncertainties. Note that, because the lower and upper uncertainties are asymmetric, the systematic biases depend on how we choose to define the default results. For instance, for the end-on galaxies, if we adopt the median as the default value, the default DM fraction at $r \lesssim 10$ kpc becomes $\sim 5\%$ lower for the model with IFU and HI, and $\sim 10\%$ lower for the model constrained by IFU only, making the bias slightly more negative correspondingly.

Previous studies have shown that the information gain in different \HI observations varies massively depending on galaxy properties and measurement quality \citep{Yasin2023MNRAS.525.5066Y}. The mock \HI spectra we created have a median peak signal-to-noise ratio of 18. Statistically speaking, the uncertainties on DM mass and DM fraction have been significantly reduced by combining \HI spectra and IFU data compared to models constrained by IFU data only for our sample. 

For comparison, we completed the full grid search using only the IFU data as model constraints for all the end-on galaxies. 
As shown in Table~\ref{tab:bias_uncertainty}, the IFU and \HI and IFU-only models provide comparable constraints in the central regions: within $2.5$--$5\,{\rm kpc}$, the relative uncertainties in the recovered DM mass are both at the level of $\sim 60$--$70\%$ with the small fraction of DM mass in this region. 
For models constrained only by IFU data, the uncertainty on DM mass becomes larger in the outer regions, which reaches $85\%$ at 20 kpc. In contrast, models constrained by IFU + \HI have smaller uncertainty on DM mass in the outer regions because DM becomes dominant and is constrained by the \HI spectrum. The relative uncertainty on DM mass at 20 kpc is $30\%$. 
The DM fraction shows a similar improvement, with the relative uncertainty at $20\,{\rm kpc}$ reduced from $^{+15}_{-33}\%$ (IFU only) to $^{+12}_{-14}\%$ (IFU + \HI). 
The corresponding one-to-one comparisons of the recovered DM fractions for the IFU-only models are shown in Appendix~\ref{app:dm_fraction_ifu_only}.

\subsection{Total and DM density profiles}

\subsubsection{Total density profile}
We show the recovery of the total density profiles in Fig.~\ref{fig:tot_density}. In both the end-on and side-on views, the recovered median profiles agree closely with the ground truth over the radial range constrained by the data, indicating that the total mass density distribution is well recovered by our modelling. 

To quantify the recovery of density slope, we defined an average logarithmic density slope over a finite radial range within our data coverage, using the enclosed masses:
\begin{equation}
\label{eqn:gamma}
\gamma_{r_1\text{--}r_2}\equiv 3 -\frac{\log \left[M(<r_2)\right]-\log \left[M(<r_1)\right]}
{\log(r_2)-\log(r_1)},
\end{equation}
where $r_1$ and $r_2$ are given in kiloparsec.

Here, we adopt $r_1 = 2$ kpc and $r_2 = 20$ kpc. 
The simulated galaxies have a median $\tilde{\gamma}_{\rm tot,\, \rm True}=1.95$, while the recovered medians are $\tilde{\gamma}_{\rm tot,\, \rm Recovered}=1.98$ for the end-on galaxies and $\tilde{\gamma}_{\rm tot,\, \rm Recovered}=2.0$ for the side-on galaxies. The total density slope defined in this region is well recovered by our model.
\subsubsection{DM density profile}
\label{sec:dm_density_profile}

We further show the recovery of the DM density profiles in Fig.~\ref{fig:dm_density}. For both the end-on and side-on galaxies, the overall DM density profiles within 20 kpc are generally recovered by our model but with some underestimation in the very inner regions ($r\lesssim 2$ kpc).

We characterised the DM density slopes using three different measures. The first was the inner density slope $\gamma$ obtained directly from the gNFW profile. The other two were $\gamma_{\rm DM,\,2\text{--}20\,{\rm kpc}}$, defined via Eq.~\ref{eqn:gamma} for the DM mass between $r_1 = 2$ kpc and $r_2 = 20$ kpc, and $\gamma_{\rm DM,\,5\text{--}20\,{\rm kpc}}$, defined analogously but between $r_1 = 5$ kpc and $r_2 = 20$ kpc.

As illustrated in Fig.~\ref{fig:dm_density}, the TNG50 galaxies have super cuspy DM inner slopes, with a median $\gamma_{\rm true} = 1.54$ obtained from direct gNFW fits across the sample of 58 simulated systems. In comparison, our models obtained $\gamma_{\rm recovered} = 1.4$ for end-on and $\gamma_{\rm recovered} = 1.2$ for side-on projections, systematically underestimating the gNFW inner slope $\gamma$. In our model, which incorporated a gNFW DM halo and stellar mass, $\gamma$ is degenerate with both the DM scale radius $r_s$ and the stellar mass-to-light ratio $M_*/L$, and is therefore not a parameter that can be robustly constrained.

By contrast, the DM density slopes $\gamma_{\rm DM,\,2\text{--}20\,{\rm kpc}}$ and $\gamma_{\rm DM,\,5\text{--}20\,{\rm kpc}}$ are better recovered. For $\gamma_{\rm DM,\,2\text{--}20\,{\rm kpc}}$, the median recovered values are $1.64$ (end-on) and $1.57$ (side-on), compared with the true median of $1.69$. For $\gamma_{\rm DM,\,5\text{--}20\,{\rm kpc}}$, the corresponding recovered medians are $1.69$ and $1.71$, relative to the true median of $1.76$. Thus, the DM density slopes measured over the radial ranges probed by the observations are reliably reproduced by our modelling and serve as robust, non-parametric tracers of the central concentration of the DM distribution.

\section{Discussion}
\label{sec:discussion}

We employed mock data generated from the TNG50 cosmological simulation to validate our method. Modern cosmological simulations such as TNG reproduce the main dynamical components---disc, bulge, bar, and gaseous disc---in a way that qualitatively resembles real galaxies, and they encompass a diversity of galaxy types that can influence dynamical measurements. Even so, these simulations remain imperfect in various respects and do not quantitatively reproduce real galaxies in all details. In the following, we outline the limitations of our mock-data-based tests for stellar kinematics and \HI spectra separately.

\subsection{Limitations of tests with simulations: Stellar kinematics}

The detailed comparison between simulations and observations shows that TNG50 galaxies have a more realistic mass--size relation than previous cosmological simulations, but they may still have some overly compact objects \citep{Pillepich2019MNRAS.490.3196P, Genel2018MNRAS.474.3976G}. The iMaNGA project has shown a significant mismatch between TNG galaxies and MaNGA regarding stellar age and metallicity distributions, along with surface mass density \citep{Manni2024MNRAS.527.6419N}. Although their internal kinematical structures are generally consistent with observations~\citep{Zhang2025A&A...699A.320Z}, TNG50 galaxies still have significantly slower or shorter bars compared to observations \citep{Frankel2022ApJ...940...61F}.
In addition, with a softening length of $\sim 0.3$ kpc \citep{Pillepich2019MNRAS.490.3196P}, stellar kinematics within three times that, i.e. $\sim 1$ kpc, may not be reliable. 

For the purpose of validating a dynamical model, we aim to uncover the ground truth of the underlying gravitational potential. We consider that the differences in stellar age and metallicities do not have a major impact here. The stabilisation of the kinematics and the unrealistic bar might affect our results. 

Of the 58 simulated galaxies used to create the mock data, 28 are classified as barred\citep{2022MNRAS.515.1524Z}, while the remaining 30 are unbarred. Although TNG50 galaxies have shorter or slower bars than real galaxies, they qualitatively capture a pattern of motion of the bar, which represents the major effects in dynamical modelling. In our model, the bar pattern motion was not explicitly accounted for; instead, the bar was represented as a thick disc composed of tube orbits. We find that when the bar-like structures are viewed end-on, our model, which does not explicitly account for bar pattern motion, still recovers the underlying mass distribution with good accuracy. The total mass is recovered well, with no significant systematic bias. However, for side-on views, the total galaxy mass can be systematically underestimated in the bar region ($r\lesssim 3$ kpc). This leads to an underestimation of either the stellar mass or the DM mass, or both, thus causing a large scatter in the DM fractions in the inner regions. In addition, an underestimation of stellar mass in the inner regions also results in a globally lower stellar mass, thus overestimating the DM mass fractions at larger radii. We expect that such bias are similar when applying the method to real barred galaxies, but the affected regions might be larger in galaxies with longer bars, given the mismatch of bar features between TNG50 and real observations \citep{Pillepich2019MNRAS.490.3196P, Roshan2021MNRAS.508..926R, Frankel2022ApJ...940...61F}.

However, many of the nominally unbarred TNG50 galaxies display bar-like, azimuth-dependent kinematic signatures in their projected kinematic maps and show similar side-on underestimation biases in the recovered total mass, especially at $r<2.5$ kpc. Thus, we considered them together with the barred galaxies when evaluating model uncertainties and systematic biases. These structures may differ from those in observed unbarred galaxies, and we expect the bias for real unbarred galaxies to be smaller if they do not exhibit such bar-like kinematic features. In addition, with the possibly unstabilised kinematics in the inner kiloparsec regions due to the softening effect, the total mass we obtained at the inner 1 kpc has a larger bias and uncertainty. Without this effect, the mass distribution in the inner kiloparsec would be better recovered by our model. 

The rotation curve uncovered by the same pipeline of the Schwarzschild code \citep{zhu2018MNRAS.473.3000Z, zhu2018NatAs...2..233Z} has been systematically compared to that observed from molecular gas (CO) for 54 CALIFA galaxies \citep{Leung2018MNRAS.477..254L}, including barred and unbarred galaxies. The circular velocities uncovered by the Schwarzschild code are consistent with CO observations within $10\%$ at $1R_e$ (about 5 kpc) and with larger scatter at $0.4R_e$, with no systematic bias. The pipeline and model construction in this paper follow exactly the same approach as~\citet{zhu2018NatAs...2..233Z}; the $\sim 10\%$ systematic bias we obtained for total mass in this paper could indeed be partially caused by the unrealistic stellar kinematics in TNG50.

\subsection{Limitations of tests with simulations: \HI distribution}

The \HI properties in the TNG simulations still show quantitative discrepancies compared to ALFALFA and xGASS observations, particularly for satellite galaxies \citep{Stenevs2019MNRAS.483.5334S}. Moreover, the spatial extent of the \HI disc derived from TNG is sensitive to the assumed atomic-to-molecular gas transition \citep{2018ApJS..238...33D}. 
Nonetheless, when the simulations are carefully mock-observed to incorporate observational systematics, TNG100 reproduces several scaling relations seen in the data, especially for gas fractions and the \HI spatial distribution at redshift zero \citep{Diemer2019MNRAS.487.1529D}. In particular, TNG100 galaxies display an \HI mass--size relation and nearly uniform radial profiles that are broadly consistent with observations \citep{2016MNRAS.460.2143W}, although the scatter in the simulated radial profiles appears to be somewhat larger than in the observations.

We selected central galaxies from TNG50 and generated the mock data using the \texttt{MARTINI} package \citep{2019MNRAS.482..821O,2024JOSS....9.6860O}, which explicitly accounts for the atomic-to-molecular transition \citep{2018ApJS..238...33D}. As discussed in Sect.~\ref{sec:hi_density}, compared to the observations, the main differences in the TNG50 mocks are that TNG50 galaxies exhibit slightly larger \HI radii and lower central densities in their normalised \HI surface-density profiles. 
To construct the \HI thin-disc model, we assumed a universal density profile that follows the median profile of the TNG50 mock sample. The scatter in this density profile can partly contribute to the scatter of DM mass we obtained, but it does not introduce a notable systematic bias. We also tested an alternative model based on the median surface-density profile derived from the observations, as presented in Appendix~\ref{app:obs_hi_density}. This choice has no appreciable impact on our final recovery of the DM mass distributions.

Our \HI disc model assumes an axisymmetric thin disc with its inclination aligned with that of the stellar component. It also assumes that the gas rotational velocity traces the circular velocity, without explicitly accounting for pressure support, warps, or lopsidedness. 
The present model is therefore most appropriate for galaxies with reasonably regular and not strongly asymmetric integrated \HI profiles, while systems with stronger asymmetries may require a more flexible treatment. \citet{2025MNRAS.539.2110Y} studied the consistency between rotation curves and integrated \HI flux profiles for SPARC galaxies and find equivalence only for galaxies passing an asymmetry screen quantified by a Bayesian information criterion (BIC)-based metric. Although we did not apply such a screening procedure in the present work, it provides a useful framework for identifying suitable systems in future applications to observational data.

\subsection{Recovery of the fitted DM parameters}

The fitted gNFW parameters are affected by degeneracy between the inner slope $\gamma$, the scale radius $r_{\rm s}$, and the stellar mass-to-light ratio. Different combinations of these parameters can produce similar enclosed mass profiles and circular-velocity curves within the observed region. Therefore, the fitted gNFW inner slope $\gamma$ should not be interpreted as a direct measurement of the asymptotic central cusp of the DM halo.

This parameter degeneracy partly explains why the fitted $\gamma$ is biased low, especially for side-on projections, while the enclosed DM mass profiles are recovered more accurately. As shown in Sect.~\ref{sec:dm_density_profile}, the average logarithmic slopes defined over radial intervals within the data coverage, such as $\gamma_{\rm DM,\,2\text{--}20\,{\rm kpc}}$ and $\gamma_{\rm DM,\,5\text{--}20\,{\rm kpc}}$, are much less affected by this degeneracy. We therefore used these radial slopes defined within the data coverage, together with enclosed DM masses and DM fractions, as our main diagnostics of the recovered dark matter distribution.

\section{Summary}
\label{sec:summary}

We developed a technique to uncover the DM distribution in galaxies by combining stellar kinematics from IFU observations and \HI gaseous kinematics represented by an integrated \HI spectrum. The stellar kinematics were modelled with a triaxial orbit-superposition Schwarzschild approach, in which the intrinsic shape parameters $(p,q)$ were treated as free parameters, allowing the inclination to vary. The \HI gas was treated as an idealised gaseous disc that shares the same inclination angle as the stellar component. Both the stellar and gas kinematics are governed by a common gravitational potential consisting of a stellar mass component, a spherical gNFW DM halo, and a central black hole. The stellar mass-to-light ratio and three DM halo parameters---the inner slope $\gamma$, the scale radius $r_s$, and the scale mass $M_{\rm DM}(<20 \,\rm kpc)$---were treated as free parameters.

 We used a sample of 58 simulated galaxies spanning a wide range of DM contents from the TNG50 cosmological simulation. These galaxies were projected at random disc inclination angles between $35^\circ$ and $90^\circ$, and for each system, we generated two mock observations with different azimuthal viewing angles, corresponding to end-on and side-on views, respectively. We generated stellar kinematics mimicking MaNGA IFU observations as well as \HI spectra analogous to those observed by the Green Bank Telescope available from HI-MaNGA. This yields a sample of 116 mock observations. We then applied our technique to the full sample, and the resulting performance in recovering the mass distribution is summarised below.

\begin{itemize}
    \item Our method successfully recovers the total, stellar, and DM mass profiles within the data coverage ($2\lesssim r \lesssim 20$ kpc). Across the full sample, the median relative systematic biases in the 2--20 kpc range are approximately $<10\%$ for the total mass, $2\%$ for the stellar mass, and $<20\%$ for the DM mass. The DM fraction is generally recovered across all radii, with a relative systematic bias of $<20\%$. For individual galaxies, the typical relative uncertainties are about $20\%$ for the total mass and $20\%$ for the stellar mass over 2--20 kpc, while the uncertainty in the DM mass decreases from $\sim 60\%$ at $r = 5$ kpc to $\sim 30\%$ at $r=20$ kpc. 

    \item By jointly using IFU and \HI data, the uncertainty in the DM mass within $\sim 5$ kpc remains comparable to that derived from models constrained solely by IFU observations, whereas the relative uncertainty in the DM mass at $r>10$ kpc is significantly reduced. At $r=20$ kpc, the uncertainty is markedly lowered from $85\%$ to $\sim 30\%$.
    
   \item The total mass density profiles between 2--20 kpc are well recovered. Moreover, the total density slope defined within these regions, $\gamma_{\rm tot,\,2\text{--}20\,{\rm kpc}}$, is well recovered for both side-on and end-on galaxies. 
   
   \item The DM density profiles within 20 kpc are also generally well recovered. The DM density slopes defined within the data coverage, $\gamma_{\rm DM,\,2\text{--}20\,{\rm kpc}}$ and $\gamma_{\rm DM,\,5\text{--}20\,{\rm kpc}}$, are reliably constrained and serve as effective indicators of the central concentration of the DM distribution. In contrast, the inner slope ($\gamma$) obtained directly from the gNFW model can be biased because of degeneracies with other model parameters.
  
\end{itemize}
 
We demonstrate that our method can robustly recover the DM mass distribution. Given that comparable IFU data and \HI spectra are available for a large number of nearby galaxies, our results provide a quantitative assessment of the expected uncertainties and systematic biases when applying this technique to real galaxies.

\begin{acknowledgement}
LZ acknowledges the support from the CAS Project for Young Scientists in Basic Research, Grant No. YSBR-062 and National Key R\&D Programme of China No. 2025YFF0511002. MY acknowledges the support of the National Natural Science Foundation of China, Young Scientists Fund project, number 12303014. This work made use of the High Performance Computing Resource in the Core 
Facility for Advanced Research Computing at the Shanghai Astronomical Observatory.
\end{acknowledgement}

\bibliographystyle{aa}  
\bibliography{TNG}

\begin{appendix}

\section{Modelling with the observed \HI density profile}
\label{app:obs_hi_density}

\begin{figure*}
\centering\includegraphics[width=0.78\linewidth]{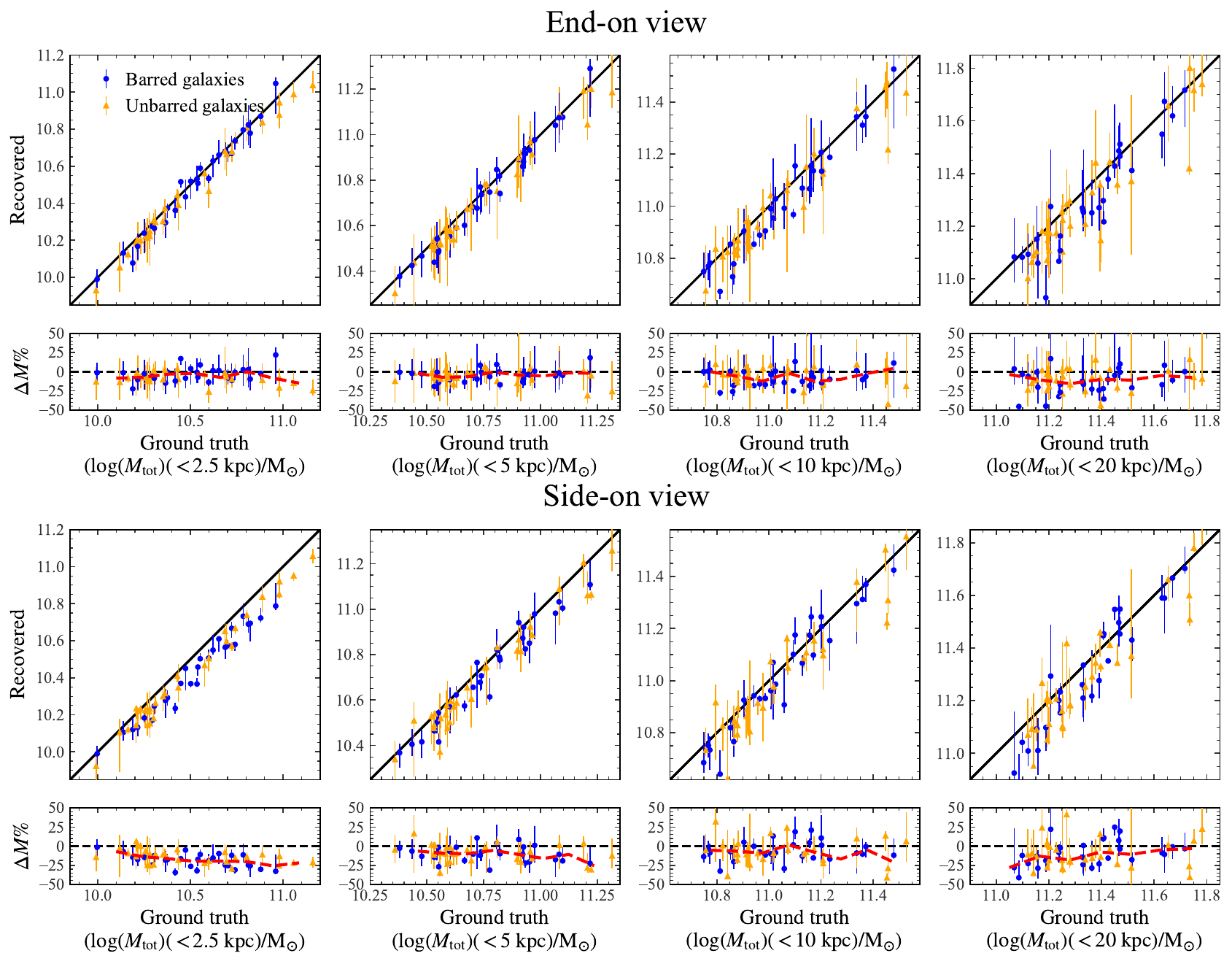}
\caption{Recovery of total mass at different radii for galaxies projected end-on.
Similar to Fig.~\ref{fig:total_mass_58} but using the alternative \HI density profile model obtained from observations.
}
\label{fig:tot_mass_new}
\end{figure*}

\begin{figure*}
\centering\includegraphics[width=0.78\linewidth]{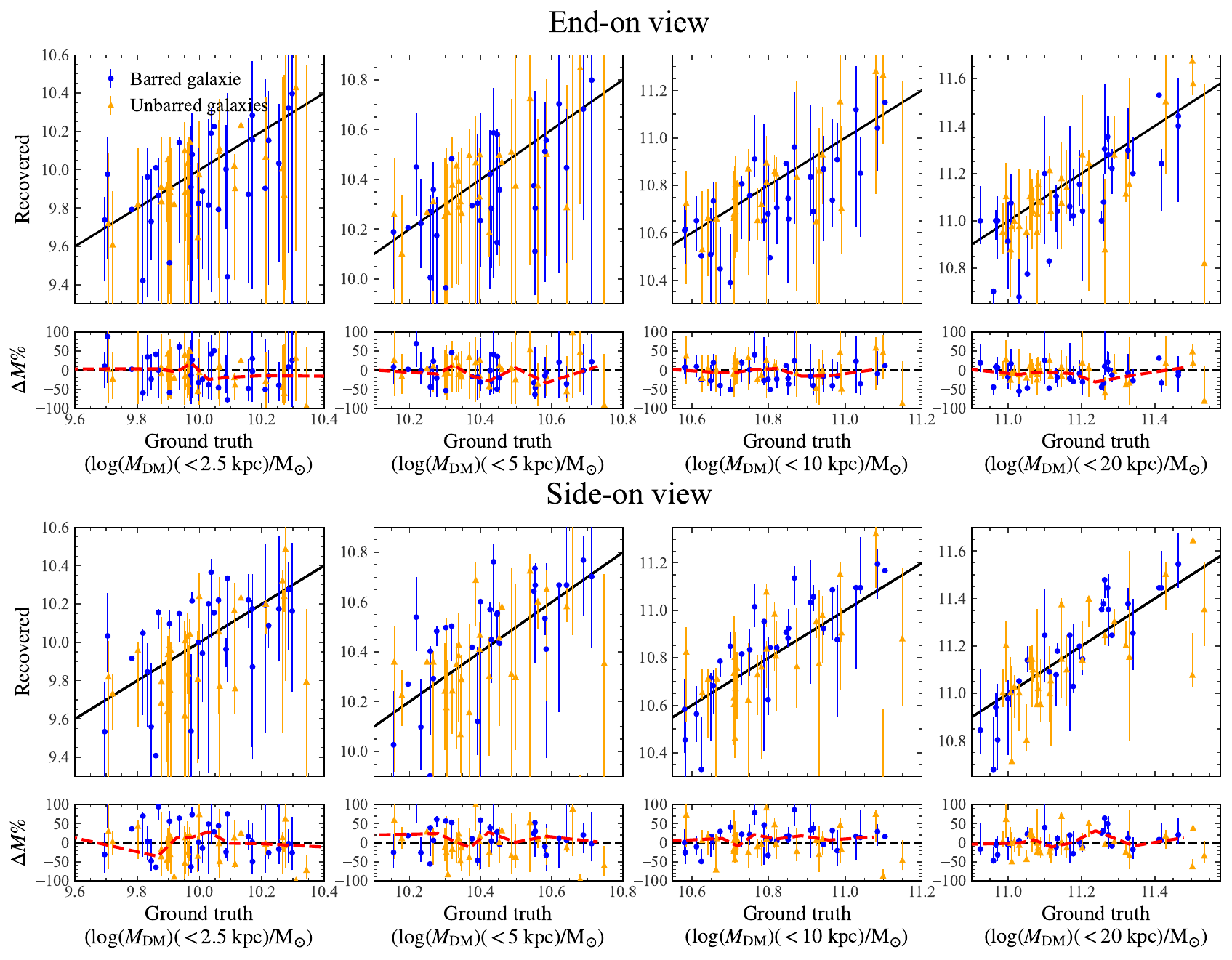}
\caption{Recovery of DM mass at different radii.
Similar to Fig.~\ref{fig:dm_mass_58} but using the alternative \HI density profile model obtained from observations.
}
\label{fig:dm_mass_new}
\end{figure*}

As discussed in Sect.~\ref{sec:hi_density}, our fiducial modelling adopts the $M_{\rm HI}$--$R_{\rm HI}$ relation and normalised \HI surface-density profile calibrated from the TNG50 mock galaxies. 
To assess the sensitivity of our results to this choice, we repeat the modelling for the mock galaxies using an alternative \HI density profile based on observations. 
Specifically, we adopt the observed $M_{\rm HI}$--$R_{\rm HI}$ relation from \citet{2016MNRAS.460.2143W} and the observed median normalised \HI surface-density profile from \citet{2020ApJ...890...63W}.
Figures ~\ref{fig:tot_mass_new} and ~\ref{fig:dm_mass_new} present one-to-one comparisons of the recovered enclosed total mass and DM mass with the ground truth for this alternative model. 
The results are similar to those obtained with the model adopting the TNG50 \HI density profile. 
This is expected because the main difference between the TNG50-based and observed \HI surface-density profiles lies in the central region, where the TNG50 profile has a lower \HI surface density. 
However, this central difference has little impact on the integrated \HI spectrum, and the inner gravitational potential is mainly constrained by the stellar kinematics. 
The enclosed total mass remains well recovered at all radii, and the enclosed DM mass is also recovered reasonably well, although it shows a comparable level of scatter at small radii. 
This test indicates that our measurements of the mass distribution are not strongly sensitive to the adopted \HI density profile.

\section{Recovery of the DM fraction and rotation curves}

\begin{figure*}
\centering\includegraphics[width=.78\textwidth]{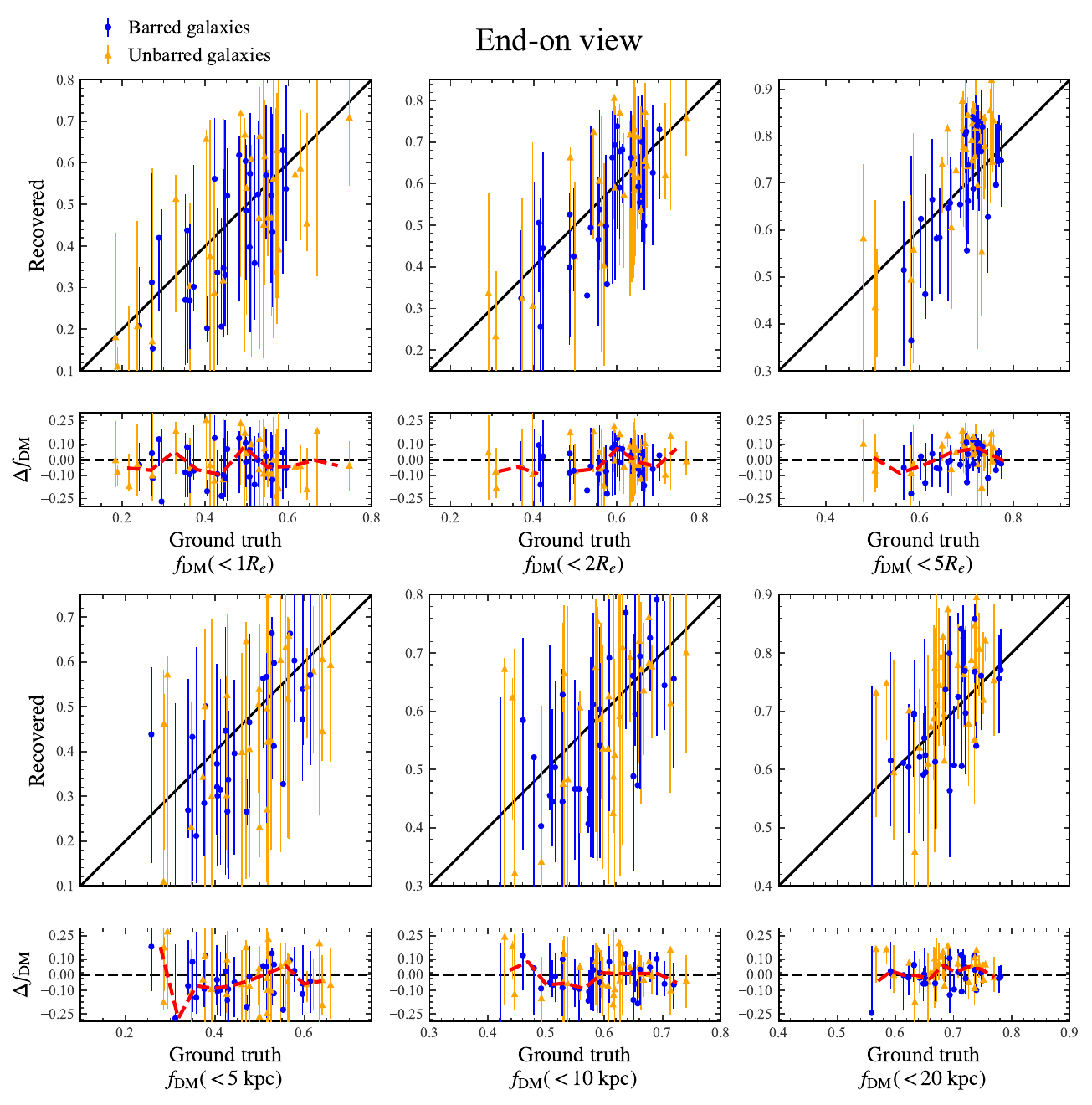}
\caption{Recovery of the DM fraction at different radii for galaxies projected end-on.
Top panel, left to right: DM fractions within $1R_e$, $2R_e$, and $5R_e$.
Bottom panel, left to right: DM fractions within 2.5, 5, and 20 kpc.
Bottom sub-panels: Deviations $\Delta f_{\rm DM} \equiv f_{\rm DM,\,recovered} - f_{\rm DM,\,true}$.}
\label{fig:dm_fraction_end_on}
\end{figure*}

\begin{figure*}
\centering\includegraphics[width=.78\textwidth]{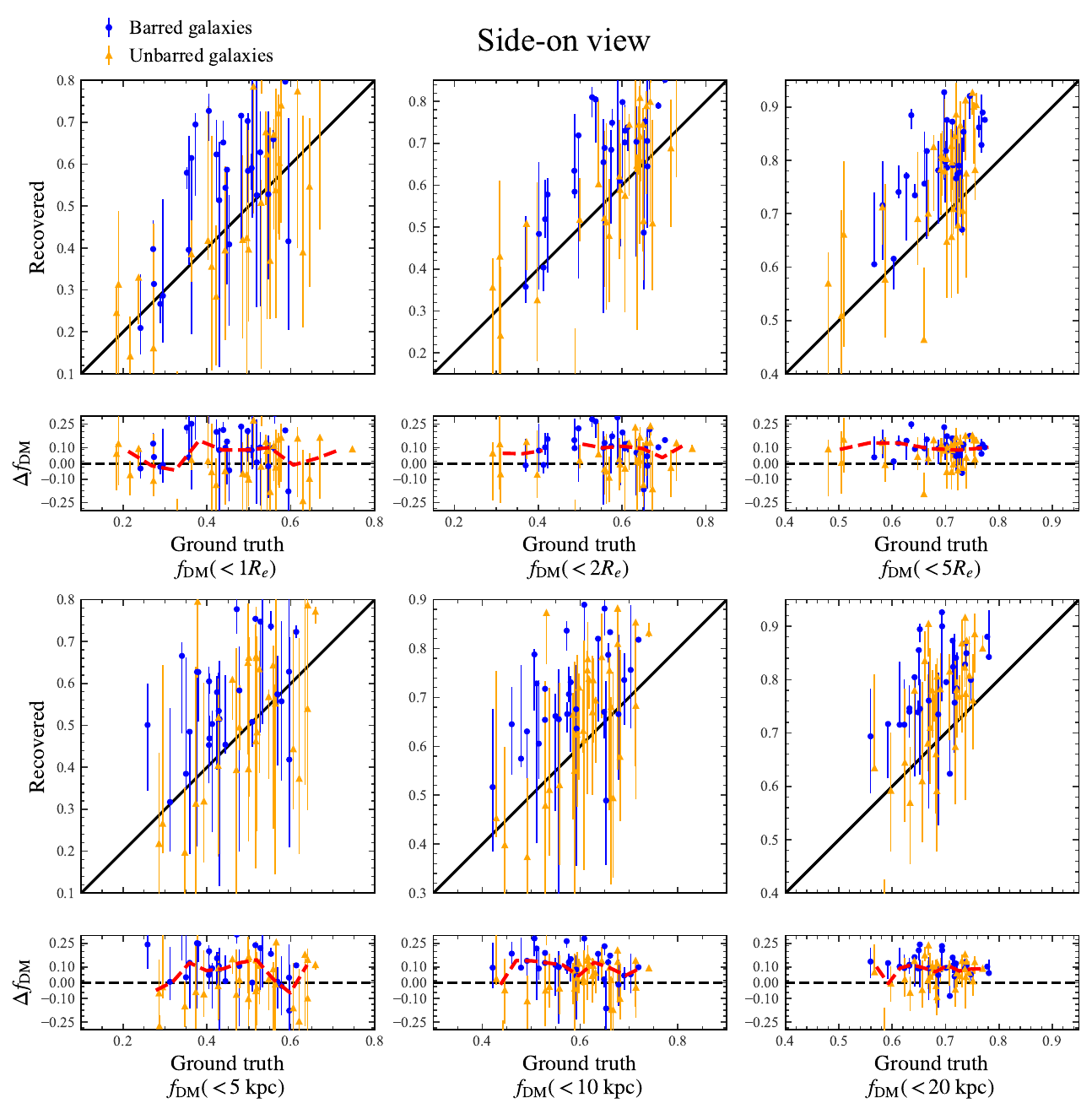}
\caption{Similar to Fig.~\ref{fig:dm_fraction_end_on} but for galaxies projected side-on.}
\label{fig:dm_fraction_side_on}
\end{figure*}

We present a one-to-one comparison of the DM fraction directly calculated from the simulations (ground truth) with that recovered by our model at several radii. As shown in Fig.~\ref{fig:dm_fraction_end_on}, the DM fraction is accurately reproduced for end-on views. For side-on views (Fig.~\ref{fig:dm_fraction_side_on}), the DM fraction is slightly overestimated, with a bias of less than $20\%$ at all radii.

\begin{figure*}
\centering\includegraphics[width=0.85\linewidth]{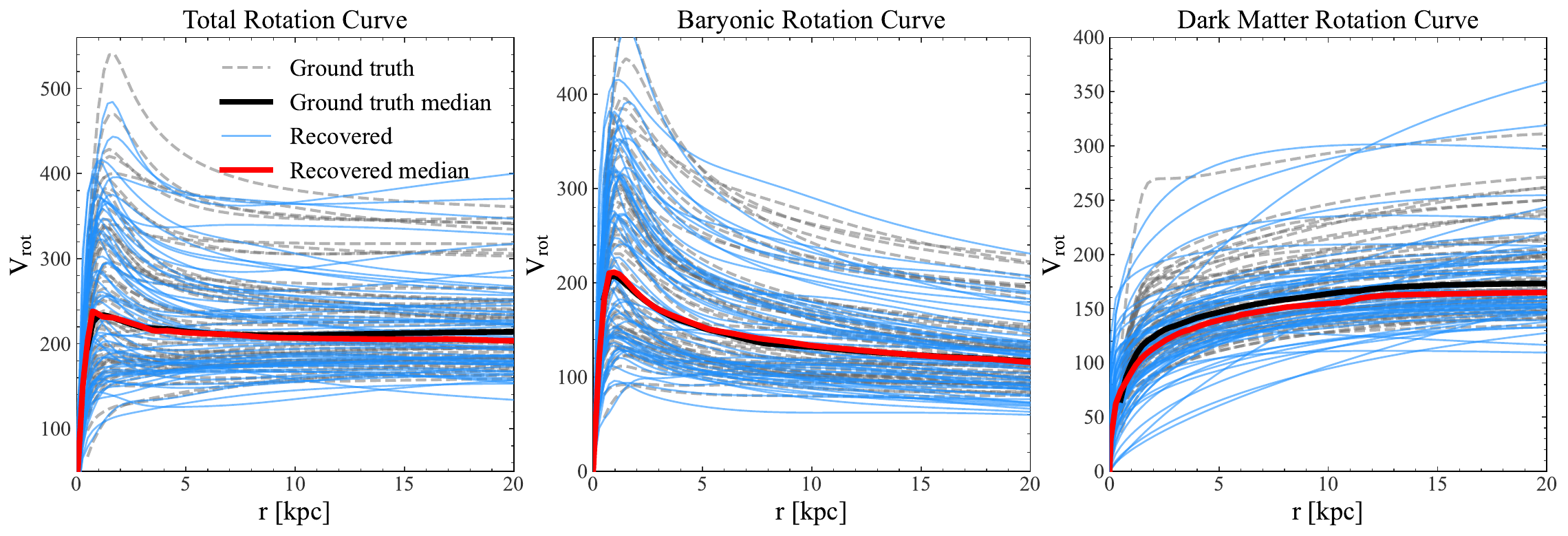}
\caption{Recovery of the rotation curve for the galaxies projected end-on. From left to right: Total, baryonic, and DM contributions to the rotation curve. Each thin grey curve represents the ground truth of one galaxy, and each thin blue curve represents the best-fitting model. The thick black and red curves represent the median of ground truth and model recovery.}
\label{fig:rot_curve_58}
\end{figure*}

The 58 simulated galaxies exhibit a diversity of rotation curves. Figure~\ref{fig:rot_curve_58} illustrates the recovery of these curves by our model, combining IFU and \HI data. The rotation curves and their contributions from baryons and DM are well reproduced, with a slight underestimation consistent with the $\sim 10\%$ bias in total and DM mass noted in Sect.~\ref{ss:totmass}. Some discrepancies appear for galaxies with extremely steep DM-dominated rotation curves, but the overall trends are reliably captured.

\section{Recovery of the DM fraction with IFU-only models}
\label{app:dm_fraction_ifu_only}

\begin{figure*}
\centering\includegraphics[width=0.78\textwidth]{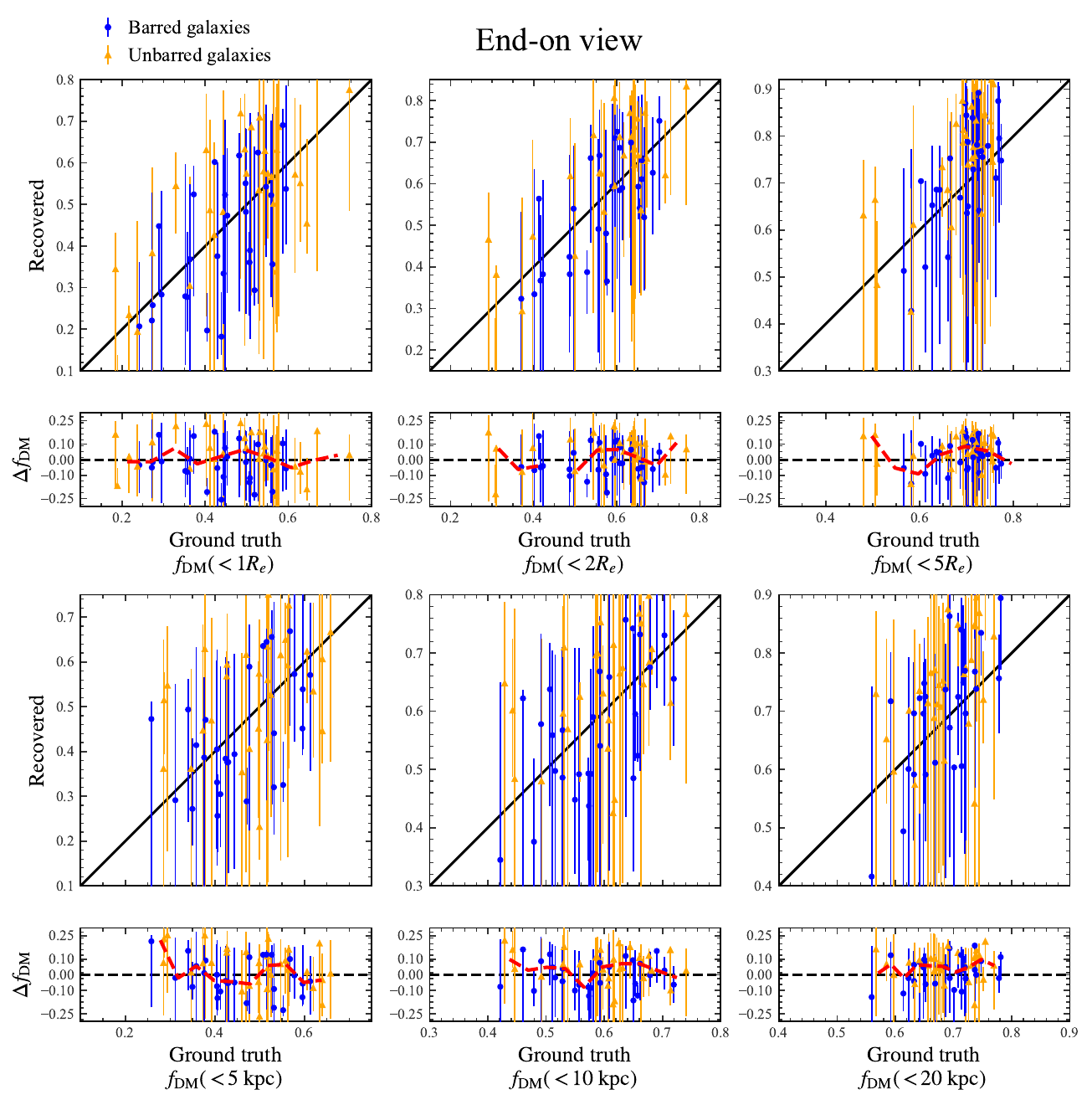}
\caption{Recovery of the DM fraction for end-on galaxies using the IFU-only model.
The symbols are the same as in Fig.~\ref{fig:dm_fraction_end_on}. }
\label{fig:dm_fraction_kinmap}
\end{figure*}

We repeat the same comparison using models constrained only by the IFU stellar kinematics.
The results are shown in Fig.~\ref{fig:dm_fraction_kinmap}. 
Compared with the IFU+\HI models in Fig.~\ref{fig:dm_fraction_end_on}, the IFU-only models recover the central DM fractions with comparable accuracy and scatter. 
This is expected because the inner gravitational potential is mainly constrained by the stellar kinematics in both cases. 
At larger radii, however, the absence of extended \HI constraints leads to substantially larger scatter, especially for the DM fractions measured at $20\,{\rm kpc}$.

\section{Azimuth-dependent kinematics for an example barred galaxy}
\label{app:azimuth_kinematics}

\begin{figure*}
\centering\includegraphics[width=0.85\textwidth]{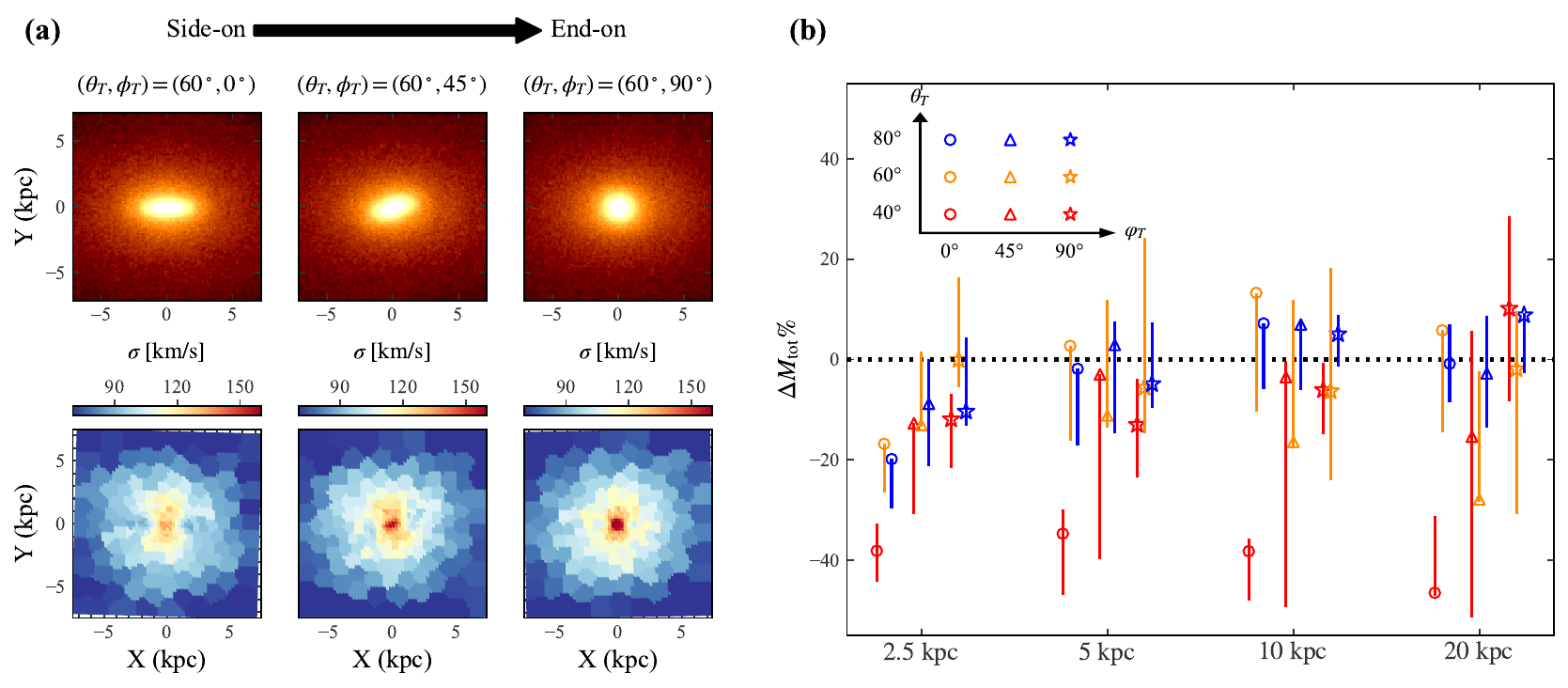}
\caption{Left: Mock stellar r-band images and velocity dispersion maps of a barred galaxy (TNG50 subhalo ID: 515695) viewed at different orientations. Top row: Mock r-band images at a fixed inclination $\theta_T=60^\circ$ and azimuthal angles $\phi_T=0^\circ$ (side-on), $45^\circ$, and $90^\circ$ (end-on) from left to right. Bottom row: Corresponding velocity dispersion maps. Right: Recovery of the enclosed total mass for the galaxy viewed at different orientations.
The figure shows the fractional deviation $\Delta M_{\rm tot}$ between the recovered and true total masses at 2.5, 5, 10, and 20 kpc.  
The symbols denote different azimuthal angles ($0^\circ$: circles, $45^\circ$: triangles, $90^\circ$: stars), while the colours represent different inclinations (
$40^\circ$: red, $60^\circ$: yellow, $80^\circ$: blue).  
A horizontal dotted line marks zero deviation for reference.  }
\label{fig:bar_ang_mock}
\end{figure*}

We present an example of the azimuth-dependent bar-like kinematic structure discussed in the main text. In the left-hand panels of Fig.~\ref{fig:bar_ang_mock}, the central velocity dispersion increases systematically as the azimuthal viewing angle changes from side-on to end-on. The right-hand panel shows the corresponding deviations in the recovered enclosed total mass: at 2.5~kpc, side-on views exhibit the largest negative offsets, whereas end-on views are generally closer to unbiased. At larger radii ($\gtrsim 5$~kpc), this azimuthal dependence becomes less pronounced, except for low-inclination side-on galaxies where negative biases remain.

\end{appendix}

\end{document}